\documentclass[11pt]{article}

\usepackage{amsfonts,amssymb,amsmath,amsthm}
\usepackage{fullpage}
\usepackage[authoryear,sectionbib,sort]{natbib}
\usepackage[utf8]{inputenc}
\usepackage{placeins}
\usepackage{titlesec}
\titleformat{\section}[block]{\Large\bfseries}{\thesection}{1em}{}
\titleformat{\subsection}[block]{\Large\itshape}{\thesubsection}{1em}{}
\titleformat{\subsubsection}[block]{\large\itshape}{\thesubsubsection}{1em}{}
\titleformat{\paragraph}[runin]{\itshape}{\theparagraph}{1em}{}[. ]
\usepackage{fancyhdr}
\usepackage{graphicx}
\usepackage{lineno}

\usepackage{amsmath}
\usepackage{color}
\DefineNamedColor{named}{Purple}{cmyk}{0.45,0.86,0,0}

\DefineNamedColor{named}{JungleGreen} {cmyk}{0.99,0,0.52,0}

\newcommand*{\bbR}{\mathbb{R}}

\title{Failure of the fittest: limits on asexual adaptation to changing environments}

\date{}

\begin{document}

\maketitle

\renewcommand{\theequation}{\arabic{equation}}
\renewcommand{\thetable}{\arabic{table}}
\renewcommand{\thesection}{\arabic{section}}
\renewcommand{\thefigure}{\arabic{figure}}
\setcounter{equation}{0}  
\setcounter{figure}{0}
\setcounter{table}{0}


\noindent{}Rebekah Hall$^{1}$, Raphaël Forien$^2$, Matthew M.~Osmond$^{3,\ast}$, Vincent Calvez$^{1,\ast}$

\noindent{}1. Univ. Brest, CNRS UMR 6205, Laboratoire de Mathematiques de Bretagne Atlantique, F-29200 Brest, France.

\noindent{}2. BioSP, INRAE, 84914, Avignon, France.

\noindent{}3. Department of Ecology \& Evolutionary Biology, University of Toronto, Toronto, Ontario, M5S 3B2 Canada.

\noindent{}$^\ast$Corresponding authors: 

mm.osmond@utoronto.ca

vincent.calvez@cnrs.fr


\bigskip


\noindent{}\textit{Keywords}: asexual reproduction; evolutionary rescue; finite-population effects

\modulolinenumbers[3]

\newpage{}

\section{Abstract}

The dynamics of adaptation in a population are informed by its mode of reproduction. Whether individuals reproduce sexually or asexually, then, can have large effects on population persistence and extinction in changing environments.
Classical theory explains this dependence via genetic variance without further accounting for a fundamentally different mode of inheritance. Recent theoretical advances capture these mechanisms, highlighting how they dramatically change the nature of adaptation in gradually changing environments. Here we show that these recent predictions, which assume infinite-population sizes, are overly optimistic about the fate of finite asexual populations when environmental change is rapid. This is because every individual in an infinite asexual population descends from a once optimal ancestor and the likelihood of finding this optimal individual in the population declines rapidly with population size. To predict the fate of finite asexual populations we derive the equilibrium phenotypic distribution assuming there exist no individuals beyond some unknown trait. As expected, we find that when the phenotypic distribution contains the optimal trait, the population's fate is consistent with previous predictions. When the optimal trait is lost, however, an additional fitness load is incurred. We further derive a prediction for the typical trajectory of an ancestral lineage, which explains the additional load: all individuals descend from a common ancestor carrying the fittest trait, which may be suboptimal. Our work uncovers a finite-population effect that is specific to asexual adaptation and distinct from genetic drift, which significantly hinders the ability of small asexual populations to persist in changing environments.

\section{Significance}

Under what conditions will a population successfully adapt and persist in the face of a changing environment? Whether a population reproduces sexually or asexually can greatly inform its adaptation, however previous mathematical models of this process often fail to capture asexual dynamics. Here, we show that because asexual populations descend from the fittest individuals, the population's success dramatically decreases when environmental change is too rapid and the fittest individuals cannot keep up with the changing optimal trait. As a result, even relatively large populations will be driven to extinction when subject to slower rates of environmental change than previously predicted.

\section{Introduction}

Sexual reproduction is widespread among eukaryotes. Why this is, however, is unclear~\citep{otto2009}. There are many risks, difficulties, and costs to sex; sexual reproduction, therefore, must provide some evolutionary advantages which counterbalance its costs~\citep{lehtonen2012,otto2009}. One way in which sexual reproduction may confer such an advantage is in adaptation~\citep{crow1965}. In the face of rapid environmental change, populations must adapt to evade extinction. To understand how and when this adaptation is successful, classic moving optimum models sought to quantify the conditions under which a population can successfully adapt in pace with a gradually changing environment, and when populations will instead be driven to extinction~\citep{lynch1991,lynch1993,burger1995,lande1996}. In a changing environment, populations are subject to a lag load: a loss in fitness due to the mean trait lagging behind the optimum trait. At rapid rates of environmental change, this lag may grow too large and drive the population to extinction. 

These models classically assume the difference between sexual and asexual populations is the amount of genetic variance in the trait that is tracking the optimum–the potentially larger amount of variance due to recombination and segregation can be an advantage in a changing environment~\citep{lynch1993,burger1999,charlesworth1993,waxman1999}. However, this does not account for the different mechanisms of reproduction which give rise to qualitative differences between sexuals and asexuals. These differences were captured by recent work taking a partial differential equation (PDE) approach which can explicitly model reproduction~\citep{garnier2023}. In sexual populations, segregation and recombination constrain the genetic variance and the magnitude of the lag is determined by selection on the mean trait. By contrast, in asexual populations, the genetic variance is unconstrained and the lag is determined by the pace at which lineages can evolve via mutation. This makes asexuals reliant on rare lineages with the fittest traits. In fact, the ancestral lineages of individuals in these models reveal that asexual populations subject to small, frequent mutations descend from a common ancestor carrying the optimal trait~\citep{calvez2022,forien2022}. 

\begin{figure}[h!]
    \centering
    \includegraphics[width=\linewidth]{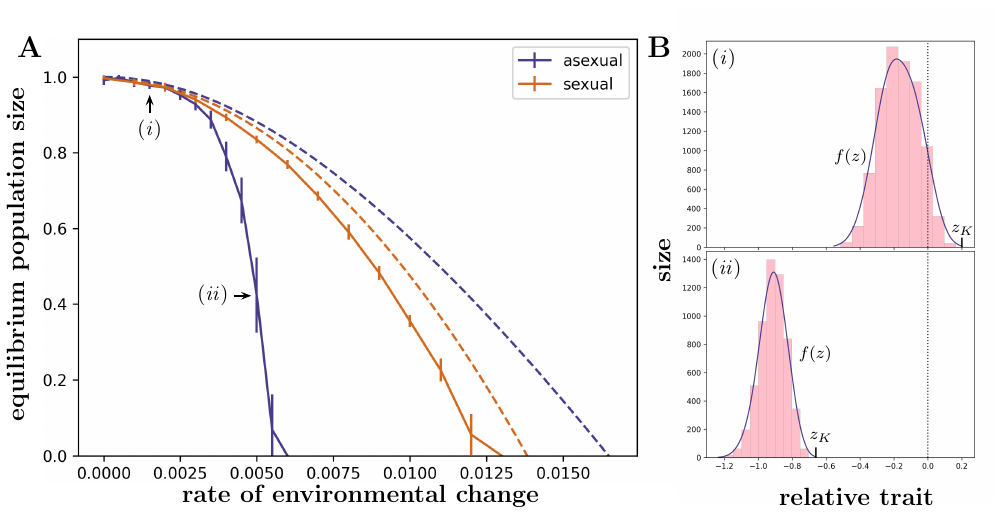}
    \caption{\textbf{Population at equilibrium.} A) Equilibrium population size of sexual (orange) and asexual (purple) populations as predicted by individual-based stochastic simulations (solid) and the infinite-population model (dotted). Simulation methods and parameters are described in Section~\ref{sec:simulations} and deterministic predictions are taken from the leading-order results presented in~\cite{garnier2023}. B) Population distribution from simulations in the asexual case, (\textit{i}) before and (\textit{ii}) after loss of the optimal trait. This distribution is modelled by a curve $f(z)$ where $z$ is the trait value relative to the optimal trait. A graphical representation of the curve $f(z)$ is traced in purple. Finite population effects are approximated by cutting this curve off at some right-most trait, $z_K$.}
    \label{fig:ace_v_allo}
\end{figure}

Yet in these models another assumption is at play: that of an infinite population. This assumption allows the inherently stochastic process of evolution to be modelled deterministically, and is assumed to accurately represent large populations in which genetic drift is weak relative to selection. However translating the model presented by~\cite{garnier2023} into an individual-based simulation reveals another discrepancy between sexual and asexual populations. While the infinite-population predictions for sexual populations match the results of finite-population simulations, this is not true for asexual populations. Simulations of asexual populations show a transition point–a speed of environmental change after which the population dynamics dramatically depart from what was predicted and the population quickly collapses as rates of environmental change increase (Fig.~\ref{fig:ace_v_allo}A). This discrepancy is dramatic even at large population sizes where genetic drift is not expected to be significant and increases as the carrying capacity is reduced (Fig.~\ref{fig:supp_ace_v_allo}), but it is not present in regimes of higher mutational variance where clonal interference is decreased (Fig.~\ref{fig:burger_sim}). This is reminiscent of dynamics in finite asexual populations purging deleterious mutations~\citep{gessler1995}, accumulating beneficial mutations~\citep{good2012,melissa2022,neher2013,roques2017,tsimring1996}, and spreading in space~\citep{brunet1997,brunet2001}, suggesting a fundamental aspect of asexuality. In particular, asexuals don't shuffle genes between lineages. This can cause limitations such as Muller's ratchet~\citep{gessler1995} and clonal interference~\citep{good2012,melissa2022,neher2013,roques2017,tsimring1996}, which depend on finite population size. Previous work has attempted to model adaptation to environmental change with clonal interference and stochastic effects~\citep{bertram2017}, but the role of population size was left unexplored.

In a moving optimum problem, assuming an infinite population supposes that all phenotypes must exist in the population at some positive frequency, which poses a problem when modelling asexual populations. While in sexual populations,  evolutionary dynamics are driven by the mean trait, dynamics in asexual populations are understood to be driven by the optimal individual in the population under the infinite-population assumption~\citep{calvez2022,forien2022}. The frequency at which the optimal trait is expected in a population may be so low that we do not expect individuals carrying this trait to be found in the population (Fig.~\ref{fig:ace_v_allo}B), similar to dynamics seen in Muller's ratchet~\citep{gessler1995}.
We anticipate that the unexpected collapse of the population is due to the loss of the optimal trait in the population. Extinction occurs when the lag load gets too large~\citep{lande1996}. However, previous approaches only consider one component of this lag load: the loss in fitness due to the movement of the optimal trait (which we call the move load). When the optimal trait is lost and no individual is optimal, there is an additional component to the lag load (which we call the loss load). We expect the loss of the optimal trait to correspond with the transition point observed in Figure~\ref{fig:ace_v_allo}.

Drawing on methods previously used to harmonise infinite-population predictions and stochastic simulations in modelling mutation accumulation~\citep{melissa2022,cohen2005,roques2017} and mutation-selection balance~\citep{champagnat2026}, we seek to modify the asexual model presented by~\cite{garnier2023} (see also~\cite{roques2020}) to allow for the possibility of losing the optimal trait. 
These models consider a population of fixed size $N$ in which the accumulation of beneficial mutations causes the population's fitness distribution to propagate forward as a travelling wave at some unknown speed $v$. As in our model, the evolutionary dynamics in finite populations diverge significantly from what is predicted by an infinite population model~\citep{tsimring1996}. While finite population effects may be formally characterised by probabilistic methods~\citep{mueller2011,roberts2021,champagnat2026}, the use of a deterministic cut-off is frequently used to capture them in a more tractable manner~\citep{brunet1997,tsimring1996,melissa2022,roques2017}. Our moving optimum problem differs from the travelling fitness wave in a key way: 
while travelling wave models fix the total population density and seek to determine the speed of the travelling wave (the rate of increase of the mean fitness), in our model the speed of the travelling wave is a fixed parameter (the rate of environmental change) and we seek to determine the total population density.
We show the loss of the optimum completely accounts for the sudden, unexpected decrease in fitness seen in Figure~\ref{fig:ace_v_allo}, demonstrate how rate of  environmental change, carrying capacity, and mutational variance determine when the optimal trait is lost, and derive an expression for the loss load.

\section{Model}

We consider a population of asexually reproducing individuals each carrying a trait $x$. The evolution of the distribution of traits $f(t,x)$ is described by a PDE consisting of a mutation term, birth term, selection through mortality, and a density-dependent mortality term (Eq.~\ref{Seq:full_pde}). The mutation term models phenotypic changes by random variation from the parent as drawn from a distribution with mean 0 and variance $\sigma^2$. Births occur at rate 1 (i.e., time is scaled by birth rate). Selection on the population is given by a quadratic mortality term $\frac{1}{2}(x-x_{opt})^2$ (i.e., trait values are scaled by selection strength). The optimal trait shifts at a constant speed $\tilde{c}=\sigma c$ so the speed of environmental change occurs on the same scale as the rate of evolution $\mathcal{O}(\sigma)$. Finally, a density-dependent mortality term enforces that the density of the total population cannot exceed 1 (i.e., population density is scaled by carrying capacity). We assume frequent, small mutations ($\sigma \ll 1$) and use the diffusion approximation (see Fig.~\ref{fig:mut_rates}). Consistent with previous moving optimum model analysis, we write our equations in terms of the trait relative to the optimum $z:=x-\tilde{c}t$. The long-term behaviour of this model is given by the stationary distribution $f(z)$, the solution to:
\begin{equation}
    0 = \frac{\sigma^2}{2} \partial_{zz}^2 f(z) + \tilde{c} \partial_z f(z) + \left (1 - \frac{z^2}{2} - \rho_\infty \right ) f(z), \qquad \rho_\infty = \int_\bbR f(z)dz. \label{eq:relative_pde}
\end{equation}
This model, which assumes an infinite population, predicts an equilibrium population density of:
\begin{equation}
    \rho_\infty = 1 - \frac{\tilde{c}^2}{2\sigma^2} - \frac{\sigma}{2}, \label{eq:lamb_inf}
\end{equation}
where the population is predicted to go extinct when $\rho_\infty$ is negative~\citep{calvez2022}. This agrees with classic results~\citep{lynch1991,lynch1993}. As in classical models, this equation for the population density displays two lag types: $\frac{\tilde{c}^2}{2\sigma^2}$ quantifies the lag load, while $\frac{\sigma}{2}$ quantifies the variance load, which under the small-variance assumption necessary for our analysis is of a smaller order. We compare this infinite population model with an individual-based simulation, in which an individual carrying relative trait $Z$ gives birth at rate 1 and dies at rate $\frac{1}{2}Z^2+\frac{N}{K}$, where $N$ is the total number of individuals and $K$ is the carrying capacity. Mutations occur with every birth and follow a normal distribution with mean 0 and variance $\sigma^2$ (see Sec.~\ref{sec:simulations} for more details).

\section{Results}

\begin{figure}[h!]
    \centering
    \includegraphics[width=\linewidth]{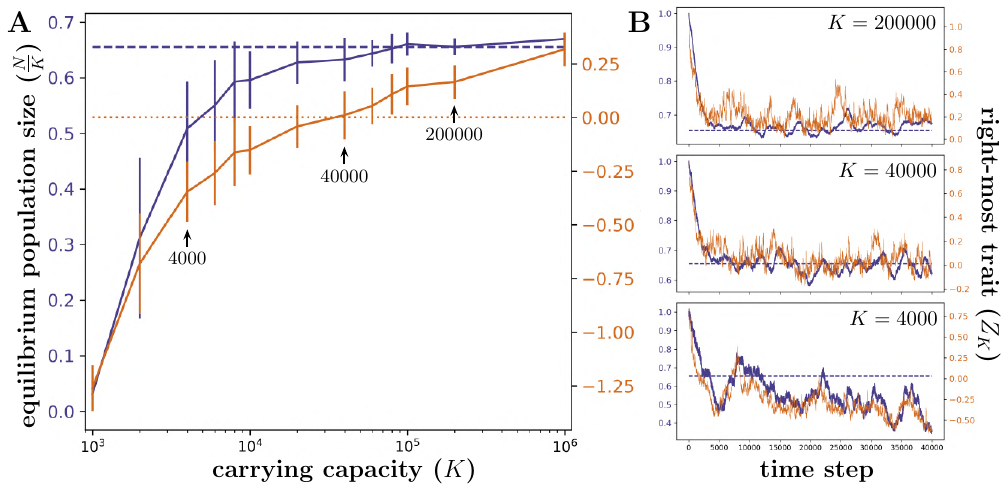}
    \caption{\textbf{Stochastic dynamics of population size (purple) and the right-most trait (orange).} (A) Population size and right-most trait at equilibrum across different carrying capacities. After a burn-in period of 10~000 time steps, simulations were run for an additional 100~000 time steps. The population size divided by the carrying capacity and the right-most trait were recorded at each time step. The mean across all time steps are plotted with error bars showing the standard deviation. The purple dashed line marks the population size predicted by Equation~\ref{eq:lamb_inf} and the orange dotted line marks $z_K=0$. (B) Dynamics of the population size relative to the carrying capacity and the right-most trait over 40~000 time steps. The purple dashed line marks the population size predicted by Equation~\ref{eq:lamb_inf}. All simulations were run for $\sigma=0.05$ and $\tilde{c}=0.04$.}
    \label{fig:logK_transition}
\end{figure}

Unlike infinite-population predictions–which are independent of any carrying capacity–the success of a population in individual-based simulations is closely tied to its carrying capacity, with observed success dropping off significantly from the prediction (Eq.~\ref{eq:lamb_inf}) at carrying capacities for which the optimal trait is lost (Fig.~\ref{fig:logK_transition}A).

We expect then that the loss of the optimal trait accounts for the sudden collapse in population size (Figs.~\ref{fig:ace_v_allo} and~\ref{fig:logK_transition}). To check this, we introduce two new elements to our deterministic (PDE) model: a carrying capacity, $K$, and a maximum (or right-most) trait, $z_K$ (Fig.~\ref{fig:ace_v_allo}B). We require that the density be zero past the right-most trait ($f(z)=0$, $z\geq z_K$) and study the equilibrium population density $\rho_K$, which we can compare to the population size relative to the carrying capacity $\frac{N}{K}$ in stochastic simulations. 

\subsection{Loss of the optimal trait}

The relationship between the right-most trait and the population size strengthens when the optimal trait is consistently absent (Fig.~\ref{fig:logK_transition}B). The optimal trait is lost when the rate of environmental change increases (Fig.~\ref{fig:ace_v_allo}) or when the carrying capacity decreases (Fig.~\ref{fig:logK_transition}). While loss of the optimal trait is a stochastic process (see Fig.~\ref{fig:logK_transition}B, middle row), we seek to characterise the conditions under which the right-most trait will on average be zero. In addition, we seek to derive a deterministic value $z_K$, which should correspond to the average value of $Z_K$ (the right-most trait measured in simulations). Figure~\ref{fig:logK_transition}A shows that when the average value of $Z_K$ is zero, the observed equilibrium population size first comes into agreement with the infinite-population prediction (Eq.~\ref{eq:lamb_inf}). To characterise this transition point, we look to the special solution of Equation~\ref{eq:relative_pde} that exists when fixing $z_K=0$, which we label $f_0(z)$. The shape of the trait distribution can be written explicitly (see Sec.~\ref{sec:supp_ctrans}) and predicts an equilibrium population density of:
\begin{equation}
    \rho_K = 1 - \frac{\tilde{c}^2}{2\sigma^2} - \frac{3\sigma}{2}. \label{eq:lamb_trans}
\end{equation}
We require the self-consistency condition that this density is equal to the density of $f_0(z)$. Following~\cite{roques2017}, we suppose that due to clonal interference, progression of the population density forward in the trait space at any given time is due to a single mutation in a single individual carrying the right-most trait (Eq.~\ref{eq:roques_condition_f0}). The size of this step forward, $\bar{s}$, is of order $\sigma$~\citep{roques2017}, however its precise value and interpretation is as yet unclear (Sec.~\ref{sec:sbar}).

In Section~\ref{sec:supp_ctrans} we use these conditions to derive the following relationship between $\tilde{c}$, $K$, and $\sigma$ at the transition:
\begin{equation}
    \ln K = \frac{\tilde{c}^2}{2\sigma^3} + \mathcal{O}\left(\ln\frac{\tilde{c}^2}{\sigma^3}\right), \label{eq:transition_pt}
\end{equation}
which shows a logarithmic dependence on $K$ (Sec.~\ref{sec:logK}; Fig.~\ref{fig:transK}). As $\tilde{c}$ increases, the minimum carrying capacity necessary to maintain the optimal trait in the population increases exponentially. As the size of mutational variation $\sigma$ increases, the minimum carrying capacity decreases. Fixing $K$ and $\sigma$, we can also numerically solve for $\tilde{c}$ in this relationship (including higher-order terms). This gives us the rate of environmental change $\tilde{c}_{trans}$ at which the optimal trait will be lost. Taking only the dominant term of Equation~\ref{eq:transition_pt}, this is approximately $\tilde{c}_{trans}\approx \sqrt{2\sigma^3\ln K}$. Comparing this with the dominant term of the critical speed in the infinite-population case, $c_*\approx\sqrt{2\sigma^2}$, we can say that the move load will drive a population to extinction before the loss load can kick in when $\tilde{c}_*<\tilde{c}_{trans}$, or when $1 \lesssim \sigma \ln K$ (Fig.~\ref{fig:sigma0_trans}).

\subsection{Population size after the loss of the optimal trait}

 We now aim to show that our analysis predicts a population size collapse when the optimal trait is lost; this loss can therefore reasonably be given as an explanation for the divergence of the stochastic simulations from infinite-population predictions. We will study the stationary distribution $f(z)$ at faster rates of environmental change $\tilde{c}>\tilde{c}_{trans}$. Our challenge is that we no longer can fix a precise value of $z_K$. We therefore cannot solve Equation~\ref{eq:relative_pde} explicitly for $f(z)$ and must take an asymptotic approach. Similarly to analysis by~\cite{cohen2005}, we consider multiple scales: when $z\ll z_K$, where population densities are high, and when $z \lesssim z_K$, where densities are low and largely driven by the right-most trait.

\subsubsection{The bulk}

The `bulk' of the population carries traits far from the the right-most trait and population densities here are large. We use the change of variables $f(z) = K^{u(z) - 1}$, where $u(z)$ is the scaling of the population density with respect to the carrying capacity (\cite{champagnat2026}; see Sec.~\ref{sec:champ}). This scaling aligns with the $\ln K$ influence on finite population effects (Fig.~\ref{fig:logK_transition} and Eq.~\ref{eq:transition_pt}; see Secs.~\ref{sec:logK} and~\ref{sec:supp_bulk}). In the limit of large $\ln K$, we can derive an explicit formula for $u(z)$, associated with a leading-order formula for $\rho_K$ (Secs.~\ref{sec:supp_bulk} and~\ref{sec:trajectories}). This is characterized by solving the cut-off equation $u(z_K)=0$, which is not an equivalent condition to $f(z_K)=0$, but performs a similar role of truncating the distribution at the right-most trait. This approximation based only on the bulk captures a large deviation in dynamics from the infinite-population case when $z_K<0$, but fails to accurately match the dynamics of the stochastic model (Fig.~\ref{fig:leading_order}). This suggests that while it is the loss of the optimal trait that drives these divergent dynamics in the finite population, a more accurate prediction requires us to determine the shape of $f(z)$ near $z_K$.

\subsubsection{The tip}

\begin{figure}[h!]
    \centering
    \includegraphics[scale=0.75]{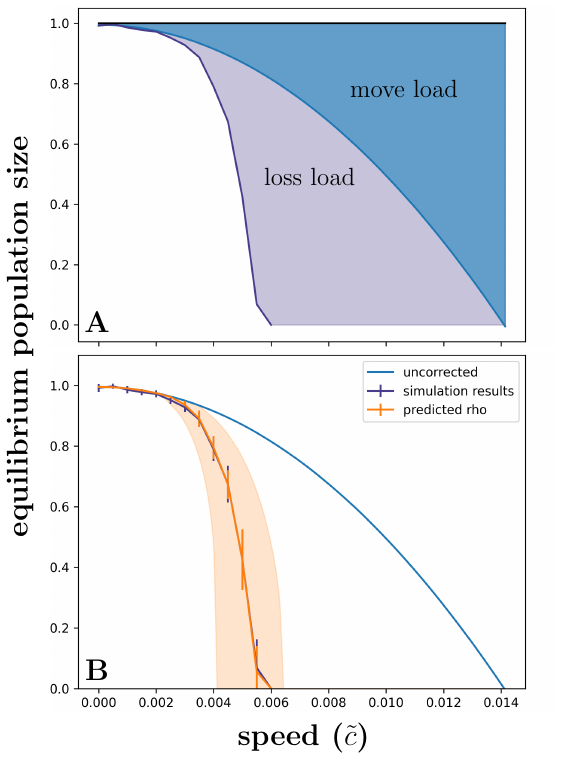}
    \caption{\textbf{Predicting the equilibrium population size by two fitness loads.} (A) Graphical representation of the primary loads determining equilibrium population size. A population may have a maximum density of 1 (black line). In blue is the fitness lost due to the changing environment (move load). In purple is the additional load due to the loss of the optimal trait (loss load). (B) Prediction of the equilibrium population size. The uncorrected deterministic prediction (Eq.~\ref{eq:lamb_inf}) is shown in blue and simulations results in purple (bars represent standard deviation over 15 replicates). In orange we plot the mean of $\rho_K$ predicted by Eq.~\ref{eq:lambdaK} plugging simulation values of $Z_K$ for $z_K$ (bars represent standard deviation over the 15 replicates; see Sec.~\ref{sec:predicting_rho}). The orange shaded region shows the range of finite population corrections with $\bar{s}=L\sigma$, where $L$ is a constant ranging from 1 to 5 (see Fig.~\ref{fig:supp_range}). Mutational variance was set at $\sigma=0.01$ and carrying capacity was $K=10~000$. Simulations were run for 15~000 time steps.}
    \label{fig:range}
\end{figure}

At the tip, population densities are lower and mutational dynamics are more significant. To better capture the collapse, we use our original equation (Eq.~\ref{eq:relative_pde}) to derive the shape of the trait distribution at the tip (Sec.~\ref{sec:supp_linearised}). This gives us a formula for the equilibrium population density as a function of the right-most trait:
\begin{equation}
    \rho_K = 1- \frac{\tilde{c}^2}{2\sigma^2} - \frac{z_K^2}{2} + \frac{\xi_0}{2}|2z_K\sigma|^{2/3}, \label{eq:lambdaK}
\end{equation}
where $\xi_0\approx -2.34$ is the first root of the Airy function (Sec.~\ref{sec:supp_linearised}). The full lag load now consists of two parts: the move load $\frac{\tilde{c}^2}{2\sigma^2}$ due to the movement of the optimal trait, and the loss load $\frac{z_K^2}{2} - \frac{\xi_0}{2}|2z_K\sigma|^{2/3}$ due to the loss of the optimal trait (Fig.~\ref{fig:range}A). Note that the move load is also the substitutional load (the difference between the fitness of the fittest trait and the mean fitness of the population), which determines the rate of adaptation~\citep{felsenstein1971,bertram2017}. When adaptation cannot keep pace with the rate of environmental change, therefore, the lag load increases while the substitutional load remains constant. 

Equation~\ref{eq:lambdaK} captures the close relationship between the position of the right-most trait in the population and the size of the population at speeds where the optimal trait is lost (Fig.~\ref{fig:logK_transition}B). When the optimal trait is lost from the population, an additional fitness load is incurred, the scale of which can be predicted by the mortality experienced by the fittest (right-most) trait in the population. If the value of the right-most trait is known from simulations, Equation~\ref{eq:lambdaK} is a near-perfect match to simulations (Figs.~\ref{fig:range}B and~\ref{fig:big_var_range}). 

\subsubsection{The right-most trait}

To predict the right-most trait $z_K$, three conditions must be met. First, at the tip we enforce the same condition as when finding the transition (Eq.~\ref{eq:roques_condition_f0}), that the progression through the trait space has an expected size of $\bar{s}$, the average size of mutations (see Sec.~\ref{sec:supp_linearised}). Next, the bulk and tip solutions must match at some match point $z_{match}$ (Sec.~\ref{sec:supp_match}). Finally, we enforce the self-consistency condition that the total density of our solution is equal to $\rho_K$ (Sec.~\ref{sec:supp_find_zk}). Having solved for $z_K$, we use this to calculate $\rho_K$ (Eq.~\ref{eq:lambdaK}).

The solution for $z_K$ captures the magnitude of the finite population correction (Figs.~\ref{fig:range}B,~\ref{fig:big_var_range},~\ref{fig:supp_range},~\ref{fig:w1correction}, and~\ref{fig:wcorrection}), however, the prediction of $z_K$ depends on $\bar{s}$: the expected progression through the trait space due to a single mutation. We see $\bar{s}$ is on the order of $\sigma$ (Fig.~\ref{fig:range}B, orange shadow), but do not know the constant of proportionality (see Sec.~\ref{sec:sbar} for an attempt). Note that due to the approximations made to maintain tractability, $z_K$ cannot be calculated for large enough values of $\tilde{c}$ (Fig.~\ref{fig:supp_range}). Given the steepness of the population collapse, however, the highest speed $\tilde{c}$ for which we can calculate $z_K$ is close to the critical speed.

\subsection{Ancestral lineages}

Following the loss of the optimal trait, the right-most trait becomes the fittest in the population. We anticipate that the close relationship between population size and the right-most trait (Figs.~\ref{fig:logK_transition} and~\ref{fig:range}B) is because the fittest takes the place of the optimal trait as the common ancestors of the population. To explore this, we use an alternative method for solving the PDE model (Eq.~\ref{eq:relative_pde}) at the bulk scale which yields predictions of the ancestral lineages (Sec.~\ref{sec:trajectories}). In an infinite population, this approach yields the leading-order approximation of $\rho_\infty$ (Eq.~\ref{eq:lamb_inf}) given by:
\begin{equation*}
    \rho_\infty^0 = 1 - \frac{\tilde{c}^2}{2\sigma^2} - \min_z\frac{z^2}{2},
\end{equation*}
where $\min_z\frac{z^2}{2}=0$ since the optimal trait, $z=0$, is always in an infinite population. This approach predicts that lineages of the current-day population will converge to a common ancestor carrying the optimal trait~\citep{forien2022}. This prediction has been confirmed by rigorous stochastic analysis of the infinite-population case [\cite{calvez2022}, generalised by~\cite{collet2024}].

\begin{figure}[h!]
    \centering
    \includegraphics[width=\linewidth]{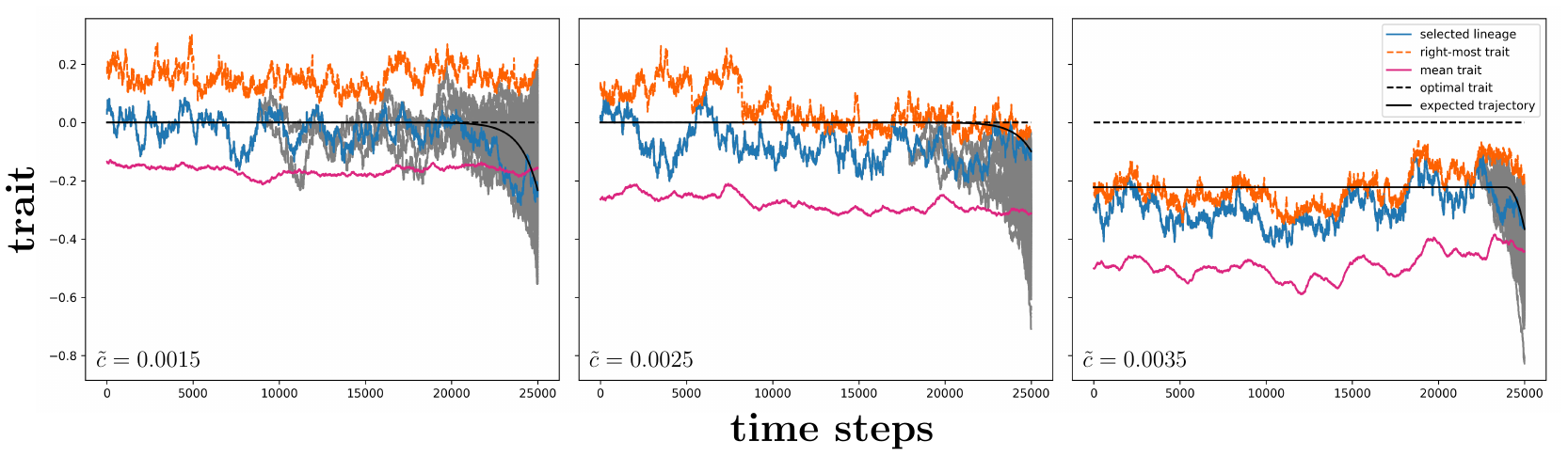}
    \caption{\textbf{Ancestral lineages of a population over the course of a moving optimum.} A random lineage is highlighted in blue and other lineages are plotted in grey. The typical lineage predicted for the highlighted lineage (Eq.~\ref{eq:typ_lineage}, using $z_K=\bar{Z}_K$) is shown in black. The right-most trait is plotted in orange, the mean trait in purple, and the black, dotted line marks the optimal trait at $z=0$. Three speeds were considered: before the optimal trait is lost (first panel), when the optimal trait is lost (second panel), and after the optimal trait is lost (third panel). Simulations were run for a burn-in period of 40~000 time steps, followed by 25~000 time steps over which lineages were recorded. All simulations were run for $\sigma = 0.01$ and $K = 10~000$.}
    \label{fig:lineages}
\end{figure}

Following the same logic in a finite population with a right-most trait $z_K$, this lineage approach yields the leading-order prediction of fitness:
\begin{equation*}
    \rho_K^0 = 1 - \frac{\tilde{c}^2}{2\sigma^2} - \min_{z\leq z_K} \frac{z^2}{2},
\end{equation*}
where the superscript 0 indicates a leading order approximation. If the optimal trait is found in the population ($z_K\geq 0$) then $\min_{z\leq z_K} \frac{z^2}{2}=0$. However, if the optimal trait is lost ($z_K<0$), the additional load term in the leading order becomes $\frac{z_K^2}{2}$ (as in Eq.~\ref{eq:lambdaK}). This perspective predicts that the ancestral lineages will converge to a common ancestor carrying the fittest trait. Given a current-day individual carrying trait $z$ and some predicted $z_K$, we derive an expected trajectory for the ancestral lineage of the individual. Where the optimal trait has not been lost, the lineages coalesce to a common ancestor around the optimal trait (Fig.~\ref{fig:lineages}, left). By contrast, when the optimal trait is lost, the lineages coalesce to a common ancestor near the right-most trait (Fig.~\ref{fig:lineages}, right). Setting $z_K$ as the average right-most trait over the course of the simulation ($\bar{Z}_K$), the expected trajectory (Sec.~\ref{sec:trajectories}) matches the observed lineage well (Fig.~\ref{fig:lineages}).

\section{Discussion}

The discrepancy between infinite-population predictions and stochastic simulations in clonal populations adapting to constant environmental change can be attributed to the low density of individuals carrying traits in the tip of the population distribution and the population's dependence on the optimal trait, from which future generations are predicted to descend~\citep{calvez2022,forien2022}. Under the infinite-population assumption, the optimal trait can always be found. In a true population, however, this trait may not be found in the population. In this case, clonal populations will suffer a fitness loss due to the loss of the optimal trait, which can lead to the observed collapse (Fig.~\ref{fig:ace_v_allo}A). We have shown that accounting for the loss of the optimal trait by introducing a cut-off in the population distribution accounts for the finite-population effects. The location of this cut-off grows exponentially further from the bulk with carrying capacity $K$ (Fig.~\ref{fig:logK_transition}A). We are able to predict both the speed of environmental change at which the optimal trait is lost and characterise the loss in fitness due to the loss of the optimal trait. Further, we show that the lineages of the modern-day population are expected to coalesce to a common ancestor carrying the fittest trait, demonstrating how the clonal population is regenerated by the fittest individuals. By contrast, in sexually reproducing populations, the population is regenerated by individuals in the bulk of the population (Fig.~\ref{fig:sex_lineages}; see also~\cite{patout2019}, Chapter 5), making the success of their adaptation less sensitive to population size effects (Fig.~\ref{fig:ace_v_allo}). 

The ability of our cut-off to predict finite-population effects is accurate only up to a certain range (Fig.~\ref{fig:range}B). This is due to a lack of information at the very tip of the distribution, around the right-most individual. While we successfully derive an accurate match to the dynamics observed in stochastic simulations (Fig.~\ref{fig:wcorrection}), our approach lacks a robust justification (see Sec.~\ref{sec:sbar}). Due to the population's dependence on the fittest trait, when the fittest trait is located in the highly stochastic tip of the distribution, small fluctuations here are expected to create large fluctuations in the bulk, as observed in other models of waves of asexual evolution~\citep{fisher2013}. Therefore, better understanding its behaviour and the relationship between the tip and the population's adaptation will require studying the underlying individual-based model and remains an open question for future work. A complete understanding of this system poses significant mathematical challenges, however our work has important connections to work on other models (see Sec.~\ref{sec:stoch_models}).

A natural comparison to our results are those of \cite{burger1999}, who found that genetic variance in asexual populations could be well-approximated by the Gaussian allelic approximation~\citep{burger2000:ch4,kimura1965} and could yield an accurate prediction for the mean population fitness across different rates of environmental change. This successful approximation arises without considering the behaviour at the tip of the population's trait distribution and holds even when the rightmost genotype is lost (Fig.~\ref{fig:burger_sim}, left). However, when applying a higher genomic mutation rate and smaller mutational variance to this simulation framework, a collapse away from the predicted population fitness is observed (Fig.~\ref{fig:burger_sim}, right). Further, the measured fitness may be retrieved by adding the loss load based on the value of the rightmost genotype. The discrepancy is therefore a question of mutational regimes. Both our work and that of \cite{burger1999} consider small, frequent mutations and fall into the regime of the Gaussian allelic approximation~\citep{burger2000:ch4}. Yet our observations demonstrate a transition within this regime. When mutations are very frequent, with very small effect size, such as in eukaryotes with large genomes, we find ourselves in the regime of pervasive clonal interference~\citep{neher2013,roques2017}. In this regime the population is subject to an additional loss load as rates of environmental change increase. As the mutation rate decreases and the mutational variance increases, we leave this regime and enter a regime where the population's response to environmental change is not well-described by Equation~\ref{eq:relative_pde}. Adaptation no longer depends on an ancestor carrying the fittest trait, but instead on a lucky lineage which undergoes successive large mutations to return to the optimum (Fig.~\ref{fig:mu_lineages}). The mutational variance is large enough for individuals from the bulk to produce offspring past the tip, thus adaptation is no longer limited by the tip. Here, the Gaussian allelic approximation is sufficient and the population is not subject to a loss load.

Our findings are consistent with theory on evolution of asexuals. We find that adaptation to a changing environment depends on the fittest individuals in the population (Eq.~\ref{eq:lambdaK} and Fig.~\ref{fig:lineages}). In genetic models, the lack of recombination means the fates of alleles are tied together, so adaptation in asexuals depends on the fittest background~\citep{neher2013,neher2013a}. This leads to so-called ``living dead'' individuals whose lineages will ultimately go extinct due to their less-fit genotype~\citep{rice2002}, hence beneficial mutations which land amongst them are lost~\citep{peck1994}. Importantly, this is only disadvantageous in finite populations, as in infinite populations there is no limitation on the fittest background~\citep{otto2009}. We also observe that as the population size decreases, the optimal trait is lost more easily and the fittest trait gets worse (Eq.~\ref{eq:transition_pt} and Fig.~\ref{fig:logK_transition}), consistent with a Muller's ratchet\textendash like mechanism. Muller's ratchet describes the repeated loss of the fittest class in a population, which cannot be reestablished in the absence of recombination or beneficial mutation~\citep{felsenstein1974,otto2021}. This is driven by genetic drift \citep{haigh1978} and by the expectation of individuals in these fitness classes being less than one~\citep{gessler1995}. The effect we capture resembles the latter, driven by expectation rather than by drift, which can be strong even in reasonably large populations. The existence of this effect in both Muller's ratchet and in our model is dependent on population size; when the force decreasing a population's fitness gets stronger (be that deleterious mutation, $Ne^{-\mu/s}<1$~\citep{gessler1995}, or changing environment, $Ke^{-\tilde{c}^2/2\sigma^3}<1$ (Eq.~\ref{eq:transition_pt})), the population size necessary to avoid this effect increases exponentially. These characteristics of asexual populations have long been offered as explanations for the evolution of sex in spite of the cost~\citep{otto2009,otto2021} and our work lends further weight to the question of how sex evolves under changing environments. 

The evolution of sex in changing environments has been previously studied in models which do not explicitly model different patterns of inheritance in sexual and asexual reproduction, nor consider dynamics at the tip. Using a classical moving optimum framework with the assumption of normality, \cite{lynch1993} showed that all else equal, recombination provides sexually reproducing populations an evolutionary advantage over asexuals. Similar predictions were made by later models~\citep{burger1999,charlesworth1993,waxman1999}. The advantage of sex was attributed to the greater variance expected in sexually reproducing populations.
For the same reason, partial clonality has also been shown to decrease a population's capacity to adapt to a continuously changing environment due to an inability to generate the constant new genetic variation necessary to track environmental change via recombination~\citep{orive2017,orive2019}.
Our results support an additional way asexuality hinders adaptation: where large effect mutations are unavailable, adaptation depends on a few key individuals which are fittest, but may not be optimal. The population is thus highly sensitive to population size (Figs.~\ref{fig:ace_v_allo} and~\ref{fig:logK_transition}). In sexual populations, however, we do not observe this same sensitivity to population size effects (Fig.~\ref{fig:ace_v_allo}). The process of segregation and recombination shift the dependence of adaptation away from a few fit individuals to the bulk of the population (Fig.~\ref{fig:sex_lineages}; see also~\cite{patout2019}, Chapter 5).

While Figure~\ref{fig:ace_v_allo} shows remarkable correspondence between previous infinite-population predictions~\citep{garnier2023} and simulations of sexual populations, this assumes an infinitesimal approximation of sexual reproduction~\citep{barton2017}. Considering only a finite number of loci is likely to lead to greater finite-population effects under sexual reproduction.
To study how genetic and demographic stochasticity shape a population's response to changing environments, \cite{burger1995} used simulations and analytical approximations of a recombining sexual population with finite loci. They found that genetic stochasticity–from factors such as genetic drift and random mutation and recombination–has large effects on the genetic variance of the population, which caused populations to go extinct quicker and at slower rates of environmental change than predicted deterministically. In sexual populations, the genetic variance of the population also changes with the rate of environmental change~\citep{burger1999}. Finite-population effects may also be more impactful on sexual populations due to inbreeding effects which arise as rates of environmental change increase and population sizes approach extinction, even in the infinitesimal model~\citep{barton2018}. Further exploration from both a deterministic and a stochastic perspective is required to better understand how finite-population effects impacts adaptation in sexual populations.

\subsection*{Acknowledgments}

The authors would like to thank Sylvie Méléard and Emmanuel Schertzer for helpful discussions on the work, and Aneil Agrawal, Denis Roze, François Ged, Puneeth Deraje, and Kuangyi Xu for their feedback on the manuscript.

VC received funding from the European Research Council (ERC) under the European Union’s Horizon 2020 research and innovation program (grant agreement No 865711) and within the the France 2030 program, Centre Henri Lebesgue ANR-11-LABX-0020-01 ; ANR grant MAMUTCELL (ANR-23-EXMA-0009).
MMO has received funding from the Natural Sciences and Engineering Research Council of Canada Discovery Grant (RGPIN2021-03207).
RF was partially supported by the Chaire ``Modélisation Mathématique de la Biodiversité'' VEOLIA-Ecole Polytechnique-MNHN-FX and the ANR project ReaCh (ANR-23-CE40-0023-01).
This project has received financial support from the CNRS through the MITI interdisciplinary programs.
RH was supported by the Natural Sciences and Engineering Research Council of Canada CGRS – D (609122-2026).

\subsection*{Conflict of Interest}

We have no conflict of interest to declare.



\FloatBarrier

\newpage{}

\bibliographystyle{abbrvnat}
\bibliography{refs}

\newpage{}

\rhead{Supplementary}
\setlength{\headsep}{0.3in}  
\lhead{} 

\setcounter{page}{1}
\setcounter{section}{0}
\renewcommand{\theequation}{S\arabic{equation}}
\renewcommand{\thetable}{S\arabic{table}}
\renewcommand{\thesection}{S\arabic{section}}
\renewcommand{\thefigure}{S\arabic{figure}}
\setcounter{equation}{0}  
\setcounter{figure}{0}
\setcounter{table}{0}

\section{Supplementary Figures}

\begin{figure}[h!]
    \centering
    \includegraphics[width=\linewidth]{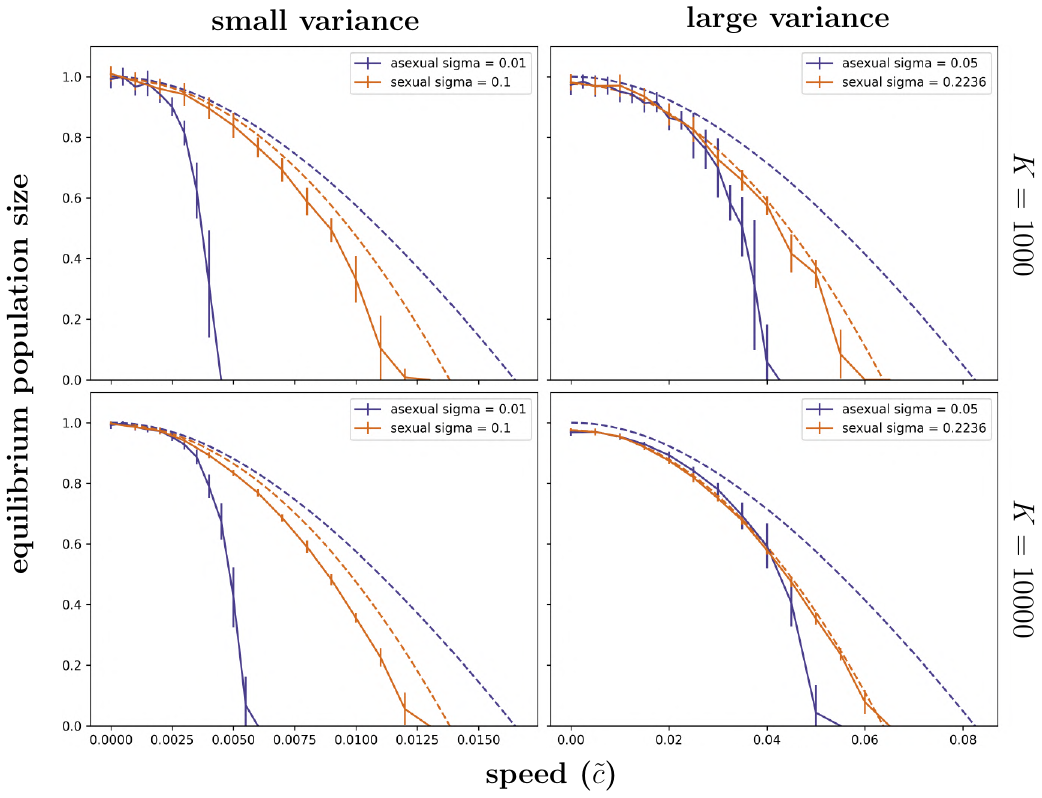}
    \caption{\textbf{Population size of sexual (orange) and asexual (purple) populations as predicted by individual-based stochastic simulations and the infinite-population model over different parameter values.} In the asexual population, mutations are drawn from a normal distribution with variance $\sigma^2$. Deterministic predictions are taken from the leading-order results presented in~\cite{garnier2023}. Deterministic predictions are plotted as dotted lines and simulation results are plotted as solid lines (bars represent standard deviation over 15 replicates). Simulations were run for 15~000 time steps.}
    \label{fig:supp_ace_v_allo}
\end{figure}

\begin{figure}[h!]
    \centering
    \includegraphics[width=\linewidth]{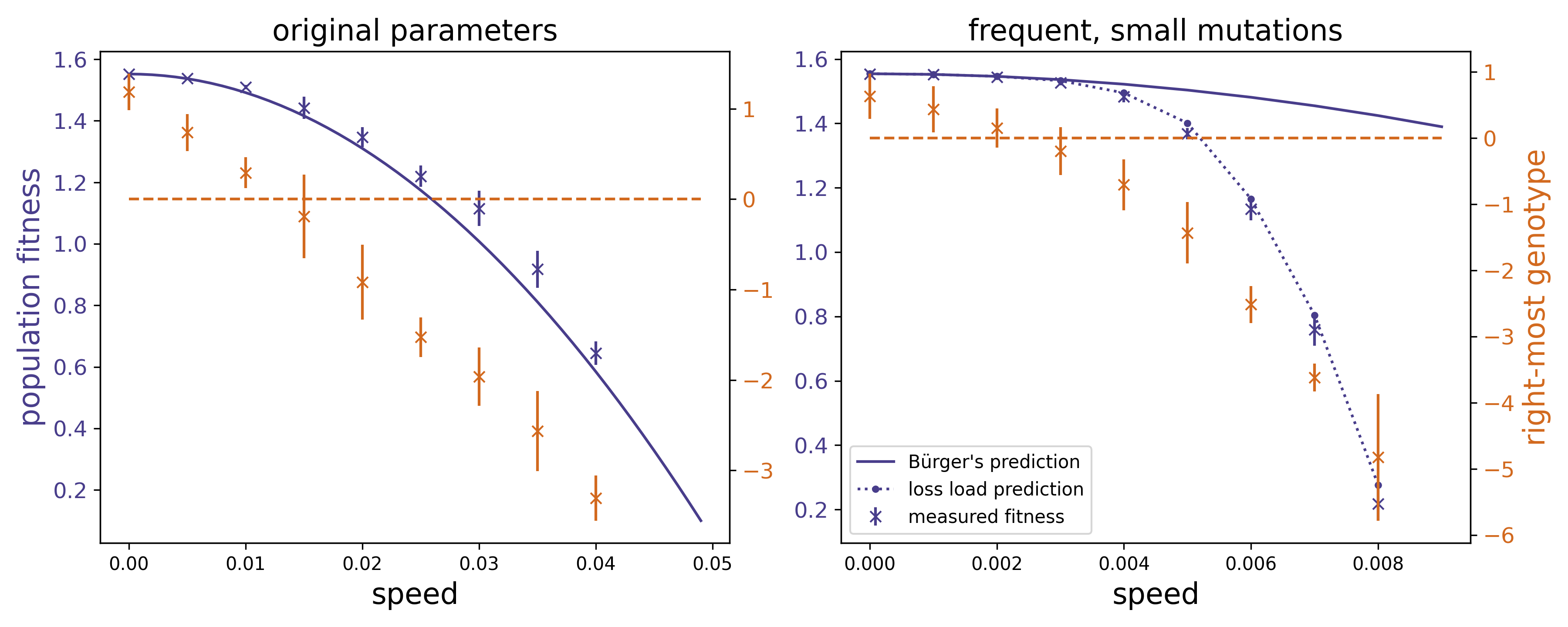}
    \caption{\textbf{Simulations of asexual adaptation under finite loci framework presented in \cite{burger1999}.} The purple line marks the prediction for asexual mean population fitness presented by~\cite{burger1999} and purple crosses mark measured mean population fitness (10 replicates, bars indicate standard deviation). Orange crosses mark the measured right-most genotype before an environmental effect is added ($Z_K$). The dotted line gives the loss load correction, multiplying the prediction from~\cite{burger1999} by $\exp{\left(-\frac{Z_K^2}{2(V_S+\sigma_a^2)}\right)}$, where $\sigma_a^2$ is the predicted genetic variance by~\citealp[Equation 10]{burger1999}. Left panel, original parameters from~\cite{burger1999}: $K=2048$, $B=5$, $\omega=3$, $U=0.02$, $\alpha^2=0.05$. Right panel, parameters chosen to reflect our simulations: $K=2048$, $B=5$, $\omega=3$, $U=0.5$, $\alpha^2=0.0005$.}
    \label{fig:burger_sim}
\end{figure}

\begin{figure}[h]
    \centering
    \includegraphics[width=0.5\textwidth]{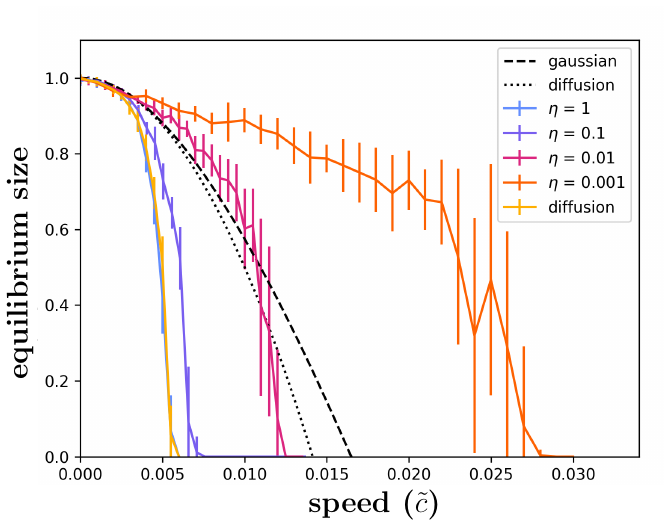}
    \caption{\textbf{Simulation results across different mutation rates ($\eta$) with a fixed $\sigma^2$.} Simulated mutations were either drawn from a Gaussian mutation kernel with variance $\sigma_{mut}^2$ at probability $\eta$ such that $\eta\sigma_{mut}^2=\sigma^2$ or were modelled by diffusion. In all cases, $\sigma$ was held fixed at 0.01. The dotted line and dashed line represent the leading-order deterministic predictions for the diffusion approximation ($\rho_{\infty}^0=1-\frac{c^2}{2}$) and the Gaussian mutation kernel from~\cite{garnier2023}, respectively. Other parameters: $K = 10000$, time steps = 15~000, replicates = 15.}
    \label{fig:mut_rates}
\end{figure}

\begin{figure}
    \centering
    \includegraphics[width=\linewidth]{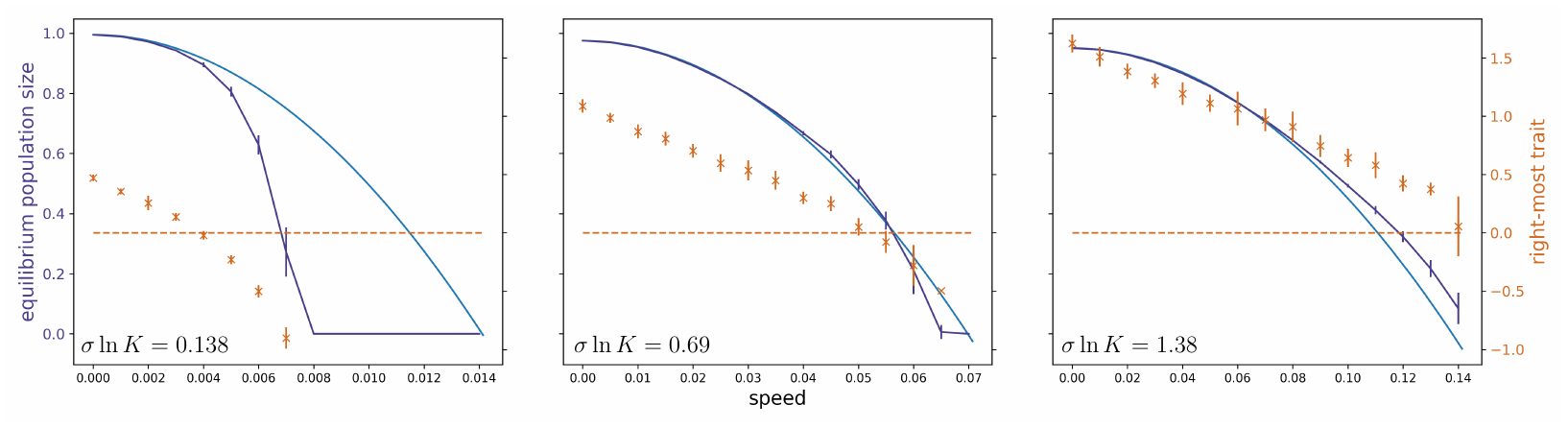}
    \caption{\textbf{Presence of loss load over different mutational variances.} Results of the population size in individual-based stochastic simulations are shown in purple (bars represent standard deviation of 15 replicates). The uncorrected deterministic prediction $\rho_\infty = 1 - \frac{c^2}{2}-\frac{\sigma}{2}$ is shown in blue. Simulations were run for 15~000 times steps with $K=1~000~000$. From left to right, the mutational variance was given by: $\sigma=0.01$, $\sigma=0.05$, and $\sigma=0.1$.}
    \label{fig:sigma0_trans}
\end{figure}

\begin{figure}[h!]
    \centering
    \includegraphics[width=\linewidth]{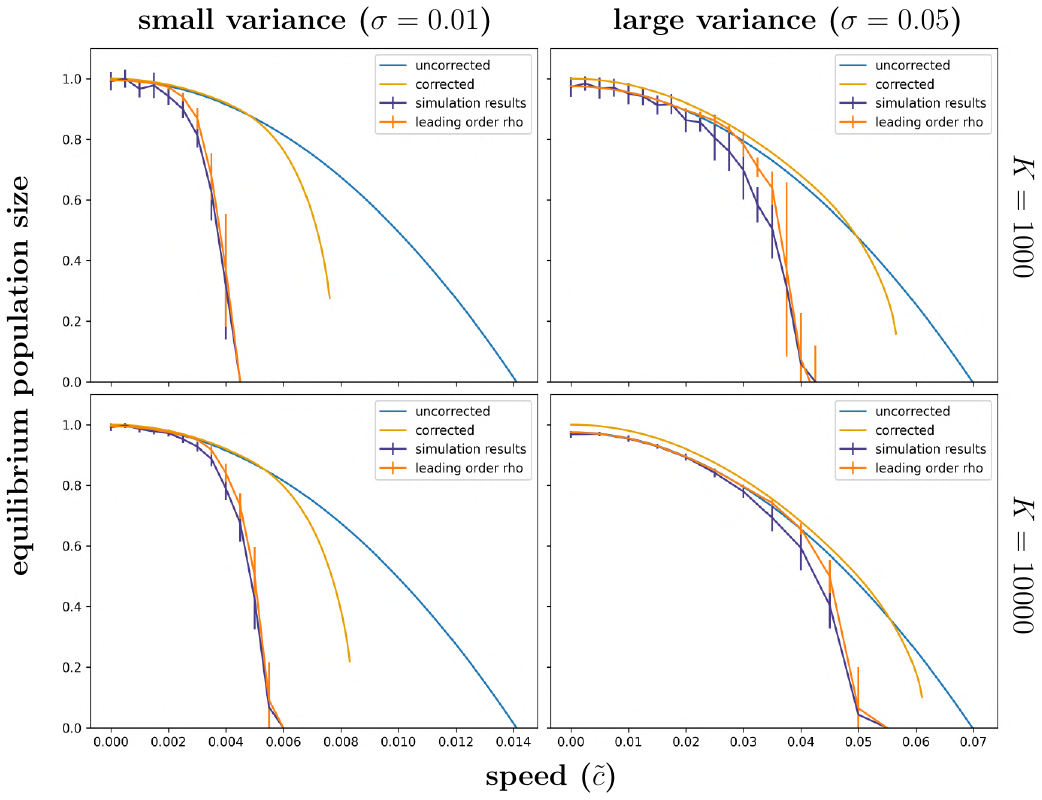}
    \caption{\textbf{Leading-order prediction of the equilibrium population size ($\rho$) according to the bulk solution with the condition $u(z_K)=0$.} Results of the population size in individual-based stochastic simulations are shown in purple (bars represented standard deviation over 15 replicates). In orange, we plot the mean of $\rho$ predicted by $1-\frac{c^2}{2}-\frac{z_K^2}{2}$ with simulation values of $Z_K$ (bars represent standard deviation over the 15 replicates, see Sec.~\ref{sec:predicting_rho}). The prediction of $\rho_K$ using $z_K$ values as predicted by the bulk solution is shown in yellow, where $\rho_K = 1 - \frac{c^2}{2}$ for speeds below the transition speed. The uncorrected deterministic prediction $\rho_\infty=1-\frac{c^2}{2}$ is shown in blue. Speed ($c$) on the x-axis is not scaled by $\sigma$. Simulations were run for 15~000 time steps.}
    \label{fig:leading_order}
\end{figure}

\begin{figure}[h!]
    \centering
    \includegraphics[width=\linewidth]{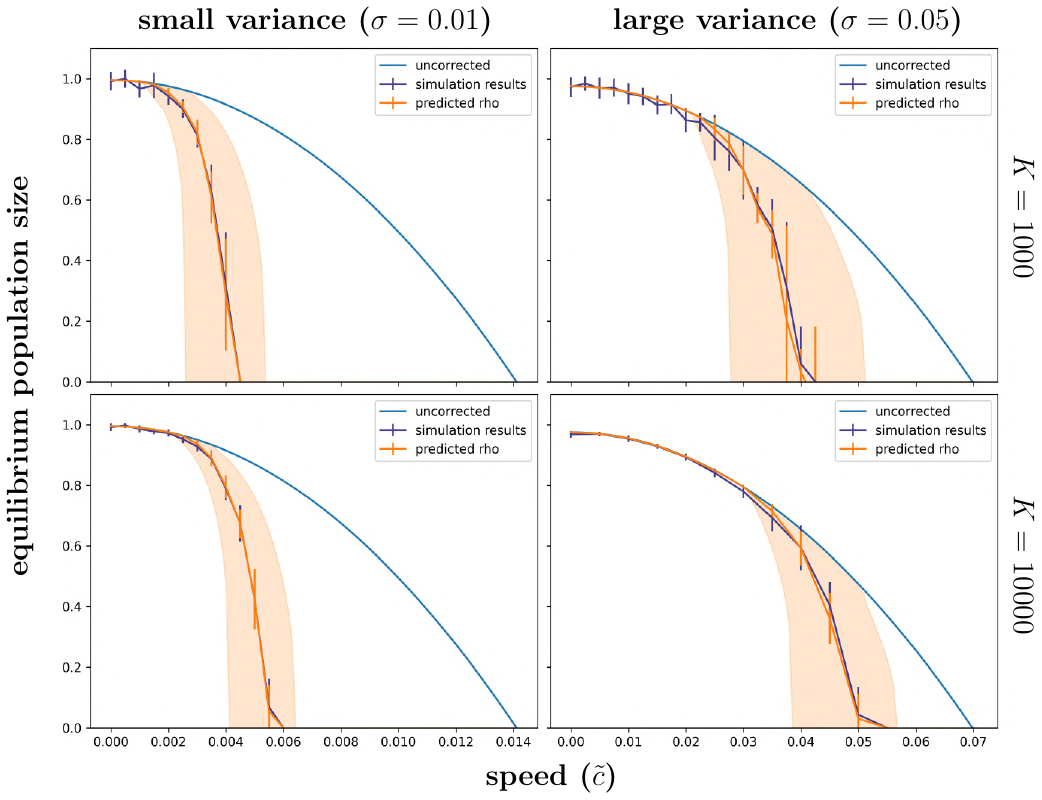}
    \caption{\textbf{Prediction of $\rho_K$ from observed $Z_K$ for different parameter values.} The uncorrected deterministic prediction (Eq.~\ref{eq:lamb_inf}) is shown in blue and simulations results in purple (bars represent standard deviation over 15 replicates). In orange we plot the mean of $\rho$ predicted by Eq.~\ref{eq:lambdaK} with simulation values of $Z_K$ (bars represent standard deviation over the 15 replicates, see Sec.~\ref{sec:predicting_rho}). The orange shaded region shows the range of finite population corrections with $\bar{s}=L\sigma$, where $L$ is a constant ranging from 1 to 5. Simulations were run for 15~000 time steps.}
    \label{fig:big_var_range}
\end{figure}

\begin{figure}[h!]
    \centering
    \includegraphics[width=\linewidth]{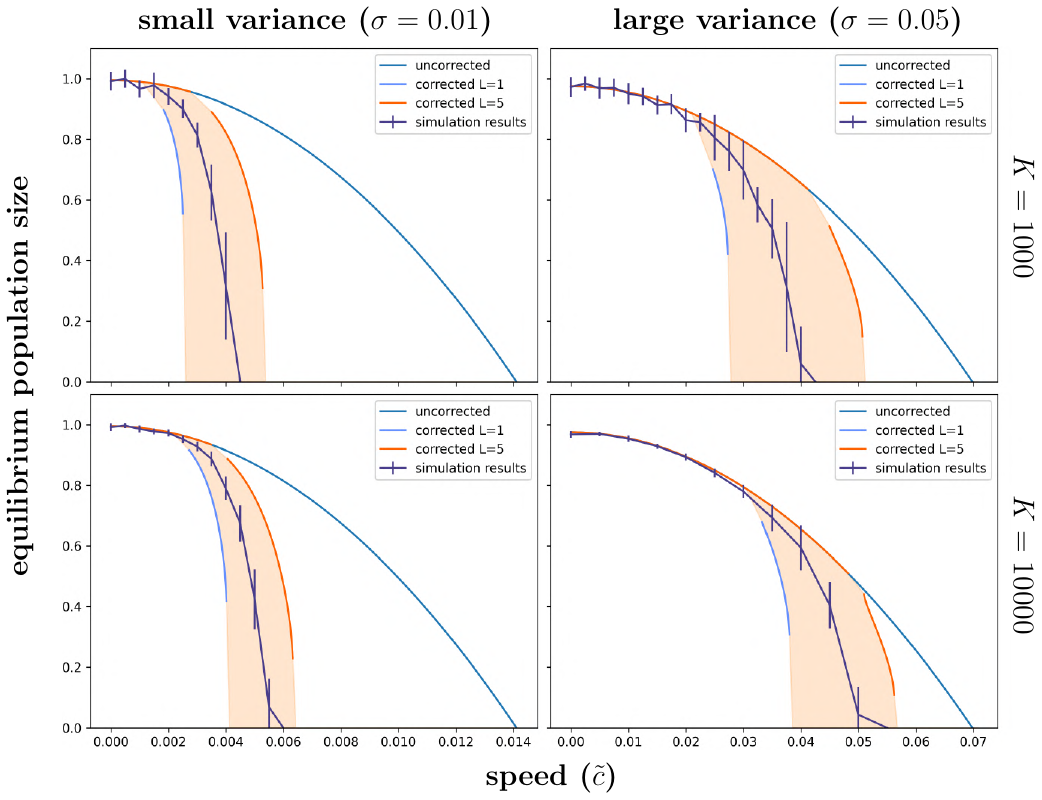}
    \caption{\textbf{Prediction of the equilibrium population size for two values of $L\sigma$.} This figure shows the corrected curves used to plot the shaded region in Figs.~\ref{fig:range}B and~\ref{fig:big_var_range}. The uncorrected deterministic prediction (Eq.~\ref{eq:lamb_inf}) is compared with the the prediction for $\rho_K$ using $z_K$ values as predicted in Sec.~\ref{sec:supp_find_zk} where $\bar{s}=L\sigma$ for $L=1$ and $L=5$. 
    Simulations results are shown in dark purple (bars represent standard deviation over 15 replicates). Simulations were run for 15~000 time steps.}
    \label{fig:supp_range}
\end{figure}

\begin{figure}[h!]
    \centering
    \includegraphics[width=\linewidth]{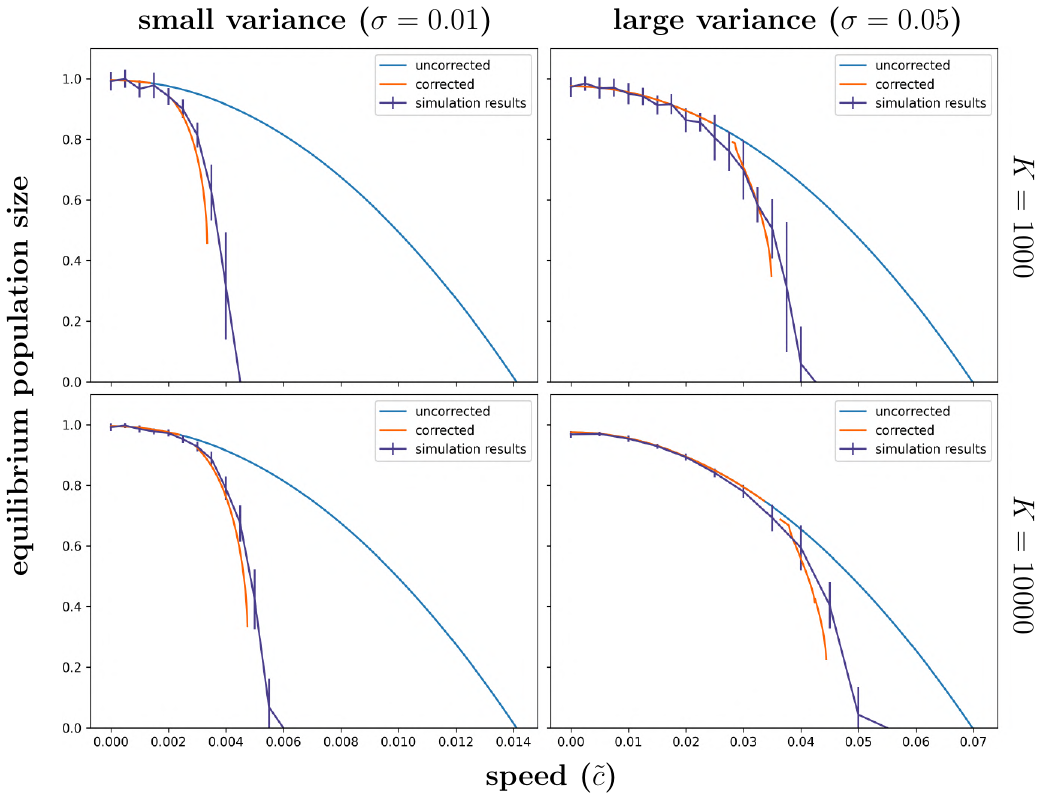}
    \caption{\textbf{Finite population correction when $\bar{s}$ is given by the expected effect size of mutations that establish with probability 1.} The uncorrected deterministic prediction of the equilibrium population size is shown in blue. In orange is the prediction for $\rho_K$ using $z_K$ values as predicted in Sec.~\ref{sec:supp_find_zk} where $\bar{s}$ is given by Eq.~\ref{eq:simp_sbar} with $w=1$. Simulation results in green (bars represent standard deviation over 15 replicates). Speed ($c$) on the x-axis is not scaled by $\sigma$. Simulations were run for 15~000 time steps.}
    \label{fig:w1correction}
\end{figure}

\begin{figure}[h!]
    \centering
    \includegraphics[width=\linewidth]{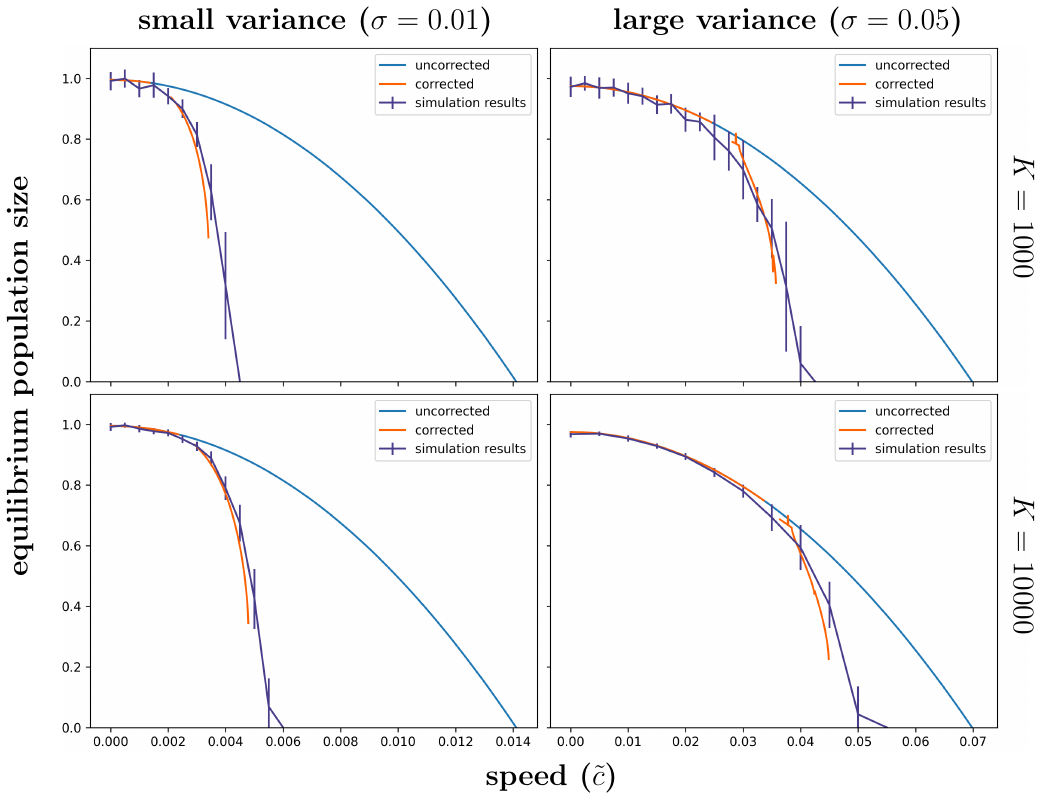}
    \caption{\textbf{Finite population correction when $\bar{s}$ is given by the expected effect size of mutations that fix under rapid mutation.} The uncorrected deterministic prediction of the equilibrium population size is shown in blue. In orange is the prediction for $\rho_K$ using $z_K$ values as predicted in Sec.~\ref{sec:supp_find_zk} where $\bar{s}$ is given by Eq.~\ref{eq:simp_sbar} with $w(z)$ as derived in Sec.~\ref{sec:find_w}. Simulation results in green (bars represent standard deviation over 15 replicates). Speed ($c$) on the x-axis is not scaled by $\sigma$. Simulations were run for 15~000 time steps.}
    \label{fig:wcorrection}
\end{figure}

\begin{figure}
    \centering
    \includegraphics[width=\linewidth]{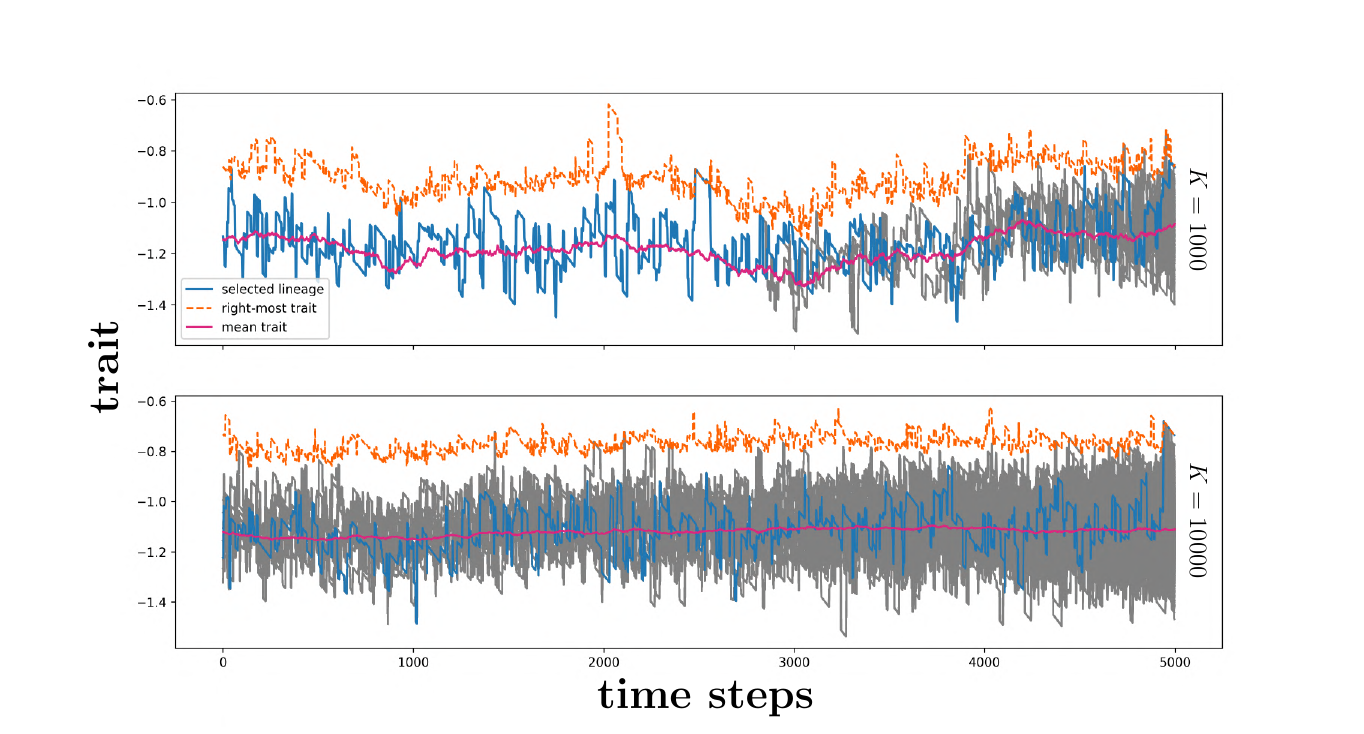}
    \caption{\textbf{Ancestral lineages of a population over the course of a moving optimum with sexual reproduction.} For each individual at each generation, the lineage shows the trait value of a randomly chosen parent. Lineages are highlighted such that for all individuals in the population a certain threshold (0.2 for $K=1000$; 0.3 for $K=10000$) above the mean trait, one lineage for each unique ancestor at time 0 is sampled. The right-most trait is plotted by a dashed purple line, the mean trait in orange, and the black, dashed line ($K=10000$) plots the mean of the highlighted lineages. Parameters: $\sigma = 0.1$, time steps = 5~000, $c = 1$, burn-in period = 40~000 time steps.}
    \label{fig:sex_lineages}
\end{figure}

\begin{figure}
    \centering
    \includegraphics[width=\linewidth]{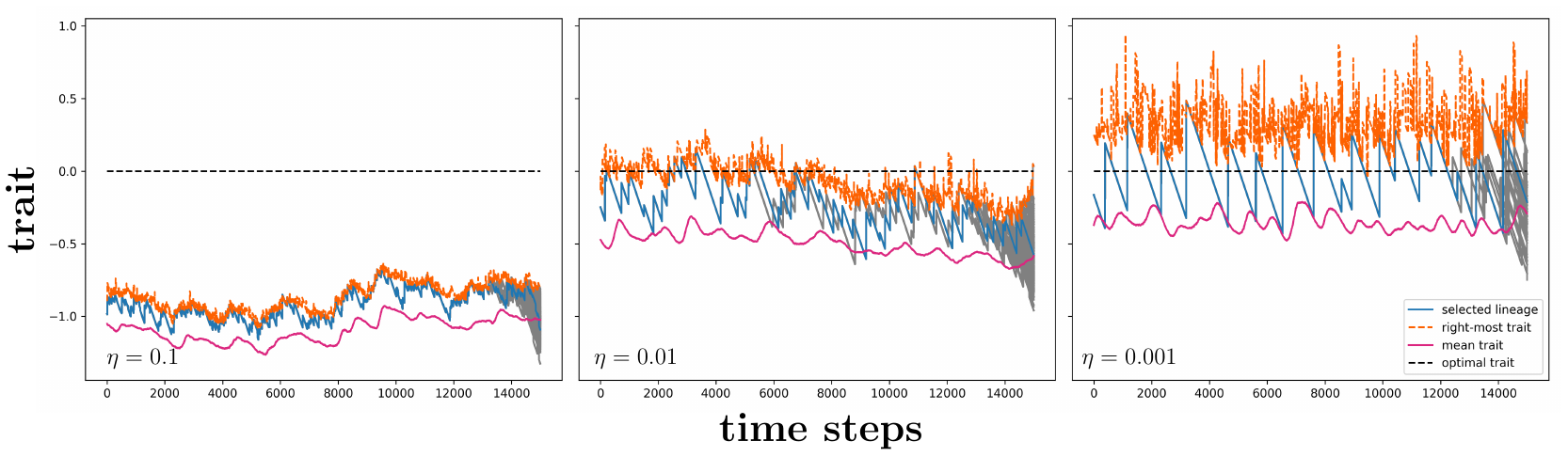}
    \caption{\textbf{Ancestral lineages of a population over the course of a moving optimum under different mutation rates ($\eta$).} A random lineage is highlighted in blue and other lineages are plotted in grey. The right-most trait is plotted in orange, the mean trait in purple, and the black, dotted line marks the optimal trait at $z=0$. Three mutation rates were considered, and $\sigma_{mut}$ was chosen such that $\eta\sigma_{mut}^2=\sigma^2=0.01^2$. Other parameters: $K = 10000$, time steps = 15~000, $c = 0.6$, burn-in period = 40~000 time steps.}
    \label{fig:mu_lineages}
\end{figure}

\begin{figure}[h!]
    \centering
    \includegraphics[scale=0.75]{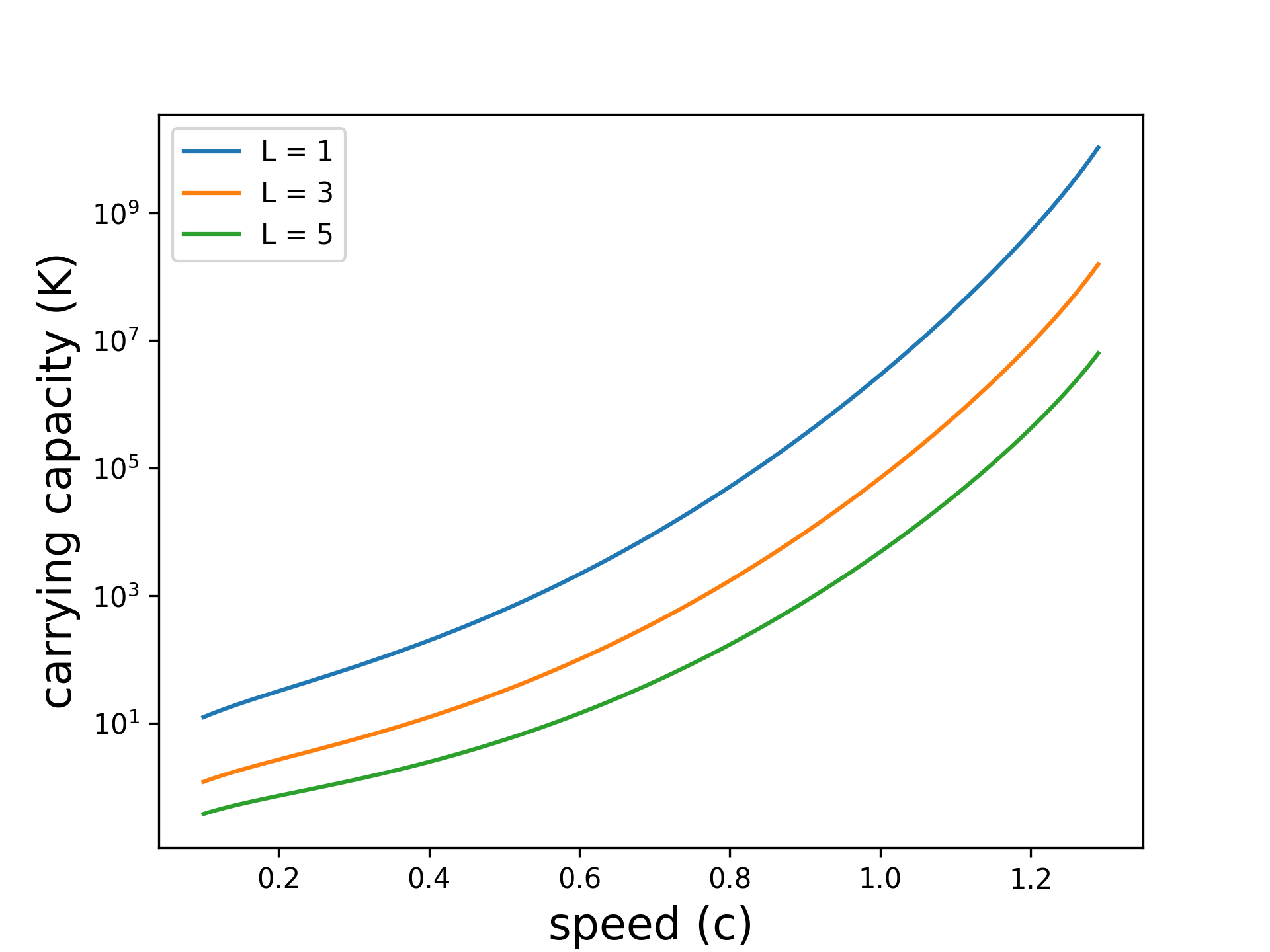}
    \caption{\textbf{Log dependence of carrying capacity ($K$).} Plot of Equation~\ref{eq:transK} with $\sigma = 0.05$.}
    \label{fig:transK}
\end{figure}

\begin{figure}[h!]
    \centering
    \includegraphics[width=\linewidth]{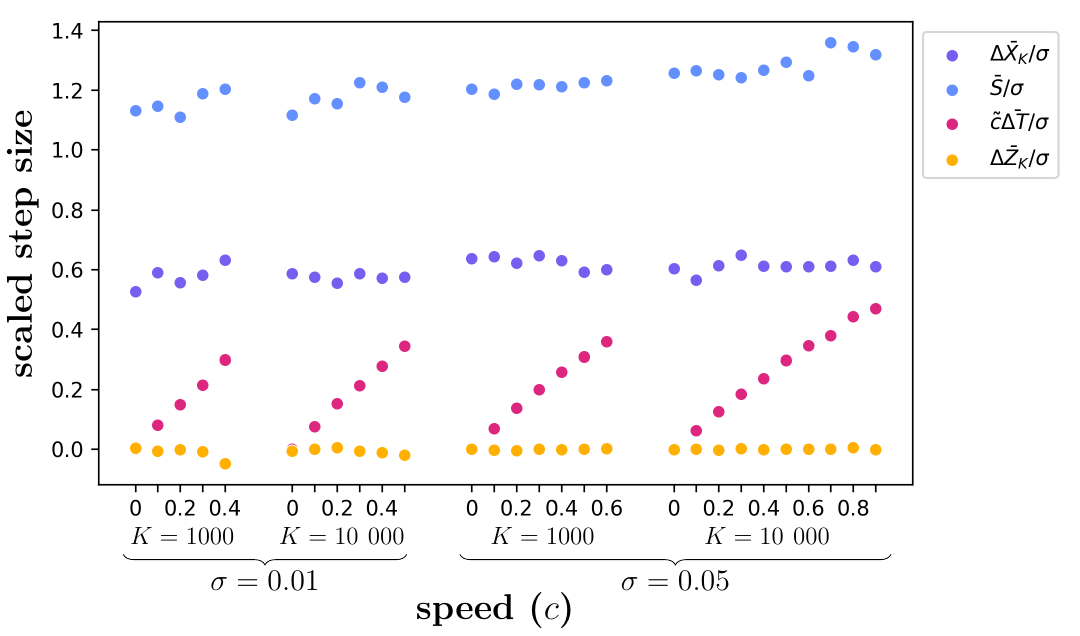}
    \caption{\textbf{Updates to the distribution with each new right-most individual.} Over the course of a simulation, whenever an offspring became the new right-most individual, we measure the following: i) the size of the step forward through trait space ($X^{t}_K-X^{t-dt}=\Delta X_K$), ii) the size of the mutation which produced the new right-most individual ($S$), iii) the size of the environmental shift since the last new right-most individual appeared ($\tilde{c}(t-t_{prev \ Z_K})=\tilde{c}\Delta T$), and iv) the change in the relative right-most trait since the last new right-most individual appeared ($Z^{t}_K-Z^{t_{prev \ Z_K}}_K=\Delta Z_K$). Measurements were taken over 1000 instances of a new right-most trait appearing, after simulations were run for a burn-in period of 10~000 time steps. Each dot shows the mean taken over these instances, scaled by dividing by $\sigma$. Results are only shown for speeds where the population did not go extinct during the initial burn-in period.}
    \label{fig:step_sizes}
\end{figure}

\begin{figure}[h]
    \centering
    \includegraphics[scale=0.75]{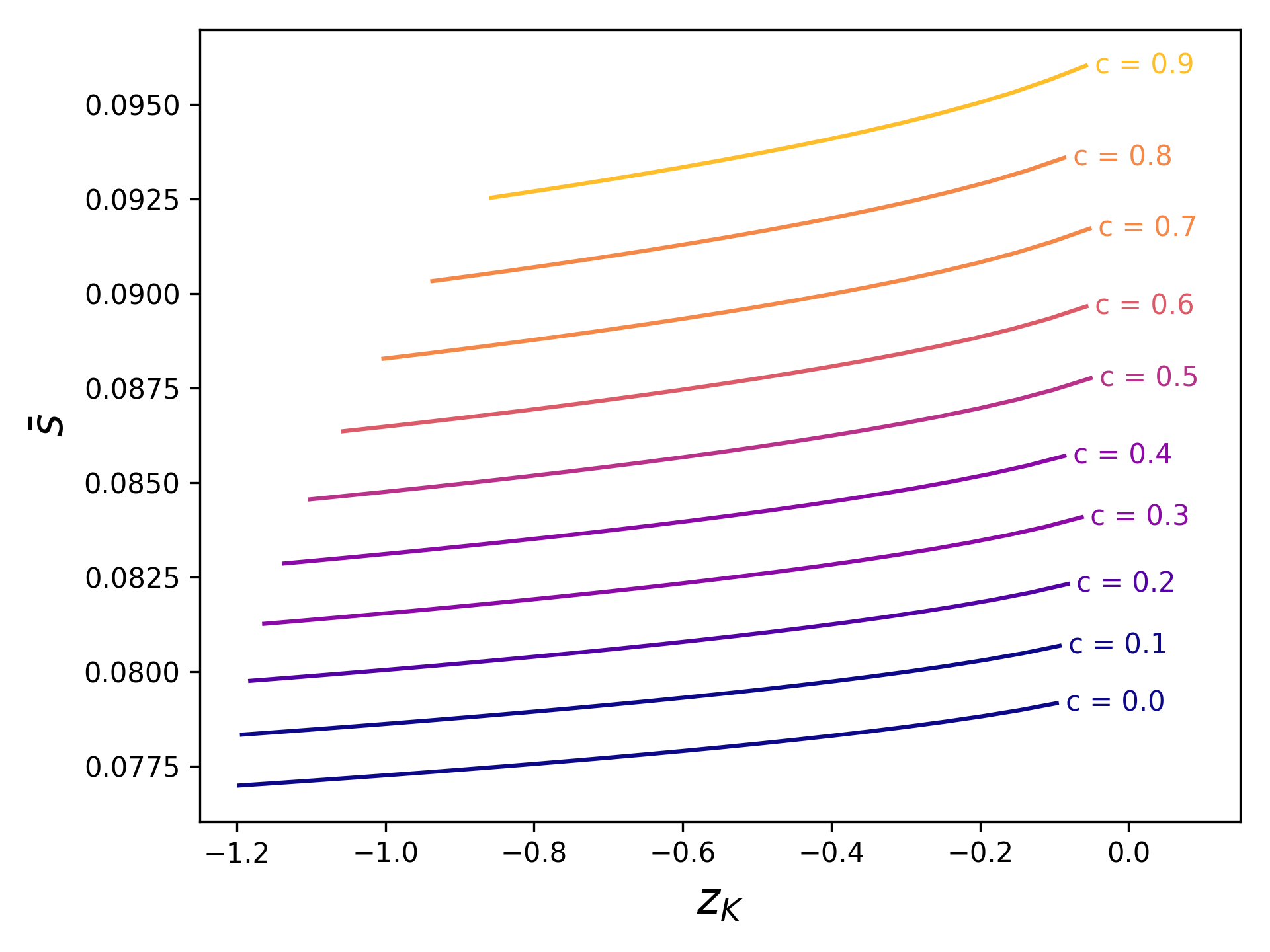}
    \caption{\textbf{Computation of $\bar{s}$ as given by Equation~\ref{eq:simp_sbar} as a function of $z_K$.} Values were computed across a range of $z_K$ values from the minimum value such that $\rho_K$ (Eq.~\ref{eq:lambdaK}) is non-negative to zero. Here $\sigma=0.05$ and $K = 10000$.}
    \label{fig:sbar_over_zk}
\end{figure}

\FloatBarrier

\section{Model} \label{sec:supp_model}

We consider the integro-differential equation:
\begin{equation}
    \partial_t f(t,x) = \frac{1}{\sigma}\int_\mathbb{R}M\left(\frac{x-x'}{\sigma}\right)f(t,x')dx' + \left (1 - \frac{1}{2}(x-\tilde{c} t)^2 - \int_\mathbb{R} f(t,y) dy \right ) f(t,x),
\end{equation} \label{Seq:full_pde}
where $f(t,x)$ denotes the density of a population at time $t$ that has trait $x \in \mathbb{R}$. Phenotypic changes are modelled by a random variation from the parent $x'$, as drawn from the distribution $M(x)$, a mutation kernel with mean 0 and variance 1. This distribution is scaled by $\sigma^2$, which represents the variance of the phenotypic changes. Mutations occur with births, and time is scaled so that the birth rate remains at 1. Selection on the population occurs through the mortality operator $\frac{1}{2}(x-\tilde{c}t)^2$ (traits are scaled by the strength of selection). Here the optimal trait is at $x_{opt}(t)=\tilde{c}t$, which shifts at a constant speed $\tilde{c}$. Individuals carrying the optimal trait experience no mortality by selection and mortality increases quadratically away from the optimal trait. We require that the speed of environmental change is on the same scale as the rate of evolution. Put otherwise, we require that $\tilde{c}$ is of order $\sigma$. To denote this in our PDE, we rewrite the speed of environmental change as $\tilde{c} = \sigma c$. Finally, the integral term $\displaystyle \int_\bbR f(t,y)dy$ models density-dependent mortality, enforcing that the density of the total population cannot exceed 1 (population density is scaled by carrying capacity).

Approximating the mutation kernel with a diffusion operator~\citep{burger2000:ch4} and shifting to the moving frame $z:=x-\tilde{c}t$, Equation~\ref{Seq:full_pde} becomes Equation~\ref{eq:relative_pde}, which has the unique positive stationary distribution:
\begin{equation}
    f(z) = \frac{\rho_\infty}{\sqrt{2\pi \sigma}} \exp \left(-\frac{(z+c)^2}{2\sigma}\right), \label{eq:inf_fz}
\end{equation}
where $\rho_\infty$ is given by Equation~\ref{eq:lamb_inf}. This equilibrium exists if and only if $\displaystyle \frac{c^2}{2} + \frac{\sigma}{2} < 1$, otherwise the population is driven extinct~\citep{calvez2022}.

To approximate the behaviour of finite populations, we introduce a new constraint on Equation~\ref{eq:relative_pde}, requiring that $f(z)=0$ for all values greater than the right-most trait $z_K$. We also rewrite the density-dependent competition term as $\rho_K$. Altogether, the finite population stationary distribution over $z\leq z_K$ can be described by the solution to the ordinary differential equation:
\begin{equation}
    \sigma^2f''(z) + 2\sigma c f'(z) + (2 - z^2 - 2\rho_K)f(z) = 0, \quad f(z_K)=0. \label{eq:ogODE}
\end{equation}
For a fixed $z_K$ value, this can be solved uniquely with the requirement that $f(z)$ is integrable and over $(-\infty,z_K]$ and strictly positive over $(-\infty,z_K)$. However, when the value of $z_K$ is unknown, a different approach is needed. 

\subsection{Identifying the transition speed} \label{sec:supp_ctrans}

Uniquely in the case of finding the transition speed, the value of $z_K$ is known. In particular, when $z_K=0$ the solution is given by:
\begin{equation}
    f_0(z)=-Az\exp\left(-\frac{(z+c_{trans})^2}{2\sigma}\right), \label{eq:f0}
\end{equation}
for $z\leq z_K$. This solution is associated with a unique $\rho_K$, which we determine by plugging Equation~\ref{eq:f0} back into Equation~\ref{eq:ogODE}. We find that the latter is only satisfied if:
\begin{equation*}
    (3\sigma + c^2 + 2\rho_K - 2)z = 0 \qquad \forall z\leq z_K,
\end{equation*}
which implies:
\begin{equation*}
    \rho_K = 1 - \frac{c_{trans}^2}{2} - \frac{3\sigma}{2}.
\end{equation*}
We note that at the leading order this is the same as what is predicted by the uncorrected deterministic model (Eq.~\ref{eq:lamb_inf}). The equation differs from the infinite-population prediction only in the variance load, $\frac{3\sigma}{2}$, which is due to the change in the shape of the phenotypic distribution.

Before determining $c_{trans}$, it remains to define $A$. This requires an additional condition. Following~\cite{roques2017}, we first suppose that at any given time, progression of the population density forward in the trait space is due to a single mutation in a single individual carrying the right-most trait. We further suppose that the expected size of this step forward is the average effect size of mutations from parent, $\bar{s}$. In this case, the constant $A$ must be chosen such that:
\begin{equation}
    K\int_{-\bar{s}}^0f_0(z)dz = 1, \label{eq:roques_condition_f0}
\end{equation}
for some $\bar{s}$. The precise value of $\bar{s}$ is as yet unclear, but of order $\sigma$. Further discussion of the calculation and interpretation of $\bar{s}$ can be found in Section~\ref{sec:sbar}. Approximating $\exp\left(-\frac{(z+c_{trans})^2}{2\sigma}\right)$ near $z=0$ with $\exp\left(-\frac{2c_{trans}z+c_{trans}^2}{2\sigma}\right)$, we find that:
\begin{equation}
    A = \frac{c_{trans}^2}{K\left (\displaystyle\left (\frac{\bar{s}}{\sigma}c_{trans}-1\right)\exp\left (\frac{\bar{s}}{\sigma}c_{trans}\right )+1\right)\sigma^2}\exp\left(\displaystyle\frac{c_{trans}^2}{2\sigma}\right). \label{eq:A_f0}
\end{equation}

Equations~\ref{eq:f0} and~\ref{eq:A_f0} together give a complete expression for the population distribution at the transition point, with the only unknown remaining the transition speed $c_{trans}$. To find $c_{trans}$, we make use of the self-consistency condition that $\rho_K$ must be the population density $\int_{-\infty}^0f_0(z)dz$. This can be solved numerically for $c_{trans}$ or approximated for $K$ as in the next section.

\subsubsection{Dependence on $\ln K$} \label{sec:logK}

Figure~\ref{fig:logK_transition} shows that the equilibrium population density depends on the carrying capacity $K$ logarithmically. To see this dependence, we can fix $\sigma$ and $c$ and derive the carrying capacity $K$ at which $z_K = 0$. To simplify this task, we write $A = \frac{1}{K}B(c,\sigma)$. Then we seek $K$ such that:
\begin{equation*}
    -\frac{1}{K}B(c,\sigma)\int_{-\infty}^0z \exp\left(-\frac{(z+c)^2}{2\sigma}\right)dz=1-\frac{c^2}{2}-\frac{3\sigma}{2}.
\end{equation*}
The left-hand integral can be well approximated by integrating over $(-\infty,\infty)$, giving $-\sqrt{2\pi\sigma c^2}$. Hence:
\begin{align}
    K &= 2\frac{B(c,\sigma)\sqrt{2\pi\sigma c^2}}{2-c^2-3\sigma} \\
    &= \frac{2c^3\sqrt{2\pi\sigma}}{\sigma^2(2-c^2-3\sigma)\left(\left(\frac{\bar{s}}{\sigma}c-1\right)\exp\left(\frac{\bar{s}}{\sigma}c\right)+1\right)}\exp\left(\frac{c^2}{2\sigma}\right) \\
    &= \left(\frac{c^2}{2\sigma}\right)^{3/2}\frac{2\sqrt{\pi}}{(2-c^2-3\sigma)((Lc-1)\exp(Lc)+1)}\exp\left(\frac{c^2}{2\sigma}\right), \label{eq:transK}
\end{align}
where $L$ is such that $\bar{s}=L\sigma$, as $\bar{s}$ should be on the order of $\sigma$ (see Section~\ref{sec:sbar}). Taking the logarithm on both sides we have:
\begin{equation*}
    \ln K = \frac{c^2}{2\sigma} + \frac{3}{2}\ln\left(\frac{c^2}{2\sigma}\right) - \ln(2-c^2-3\sigma) - \ln((Lc -1)\exp(Lc)+1) + \ln(2\sqrt{\pi}).
\end{equation*}
Assuming $\sigma$ small, $\ln K$ can therefore be written as a term of order $\mathcal{O}\left(\frac{c^2}{\sigma}\right)$, followed by a term of order $\mathcal{O}\left(\ln\frac{c^2}{\sigma}\right)$, followed by a term of order $\mathcal{O}(1)$. Clearly then, the carrying capacity has a logarithmic effect on when the optimal trait is lost. This log dependence can be seen graphically in Figure~\ref{fig:transK}.

\subsection{Characterising population dynamics past the transition speed}

\subsubsection{Solution in the bulk} \label{sec:supp_bulk}

Past the transition point, the optimal trait is lost; we thus assume $z_K<0$ going forward. We first consider Equation~\ref{eq:ogODE} when $z\ll z_K$, where the majority of the population is to be found. We call this the `bulk' of the population and expect a large population density here. To find a bulk solution we perform a modified Cole-Hopf change of variables (\cite{champagnat2026}; see Sec.~\ref{sec:champ}):
\begin{equation*}
    f(z) = K^{u(z) - 1} = \exp \left(\ln K \ (u(z) - 1) \right).
\end{equation*}
The traditional Cole-Hopf transformation has already been used in the deterministic case to capture the fact that taking the mutational variance to zero, $\sigma \to 0$, the population will concentrate around a single trait~\citep{barles2009,diekmann2005,perthame2007,garnier2023,forien2022}. By introducing a base of $K$ rather than $e$, we can describe the population density as it is scaled with respect to the carrying capacity. Doing so in the infinite-population case, Equation~\ref{eq:inf_fz} can be written with:
\begin{equation*}
    u(z) = -\frac{(z+c)^2}{2\sigma\ln K} + C,
\end{equation*}
which naturally gives the scaling $\sigma_0=\sigma\ln K$. This same scaling arises when putting the exponential part of Equation~\ref{eq:f0} to a base of $K$ and in condition for the loss of the optimal trait (Eq.~\ref{eq:transition_pt} where $\tilde{c}=\sigma c$, and Sec.~\ref{sec:logK}). In our asymptotic analysis, we imagine $\sigma$ to approach zero and $\ln K$ to approach infinity. The value $\sigma_0$ represents the balance between these two limits. This allows us to consider a range of parameter values within the case of large population sizes ($\ln K \gg 1$) and small phenotypic variances ($\sigma\ll 1$), in which these finite population effects arise (Fig.~\ref{fig:sigma0_trans}). We note that the scaling $\sigma\ln K=1$ appears in the rigorous characterisation of finite-population effects in a related mutation-selection balance model (\cite{champagnat2026}; see Sec.~\ref{sec:champ}).

In the case of Equation~\ref{eq:ogODE}, from this rescaling we arrive at the ODE:
\begin{equation*}
    -\sigma c \left (\ln K \right)u'(z) = \frac{\sigma^2}{2}\left ( \ln K \right)^2 \left(u'(z)\right )^2 + \frac{\sigma^2}{2}\left (\ln K \right ) u''(z) + 1 - \frac{z^2}{2} - \rho_K.
\end{equation*}
Letting $\varepsilon = \frac{1}{\ln K}$, and multiplying through by 2, we get:
\begin{equation*}
    \varepsilon\sigma_0^2u''+\sigma_0^2(u')^2 + 2c\sigma_0u' + 2 - z^2-2\rho_K = 0,
\end{equation*}
which gives the leading order equation:
\begin{equation}
    \sigma_0^2(u')^2 + 2c\sigma_0u' + 2 - z^2-2\rho_K = 0. \label{eq:bulk_ode}
\end{equation}
This change of variables and scaling allows us to track only the large deviations caused by mutations that drive the distribution ahead, while dropping smaller deviations to the lower order. 

We start by attempting to solve this problem on only the scale of the bulk. In this case, our initial condition of $f(z_K)=0$ cannot apply. We instead take $u(z_K)=0$, so that $f(z_K)$ is on the order of $K^{-1}$, corresponding to few individuals. Following a similar argument to that presented by~\cite{garnier2023}, we can complete the square to rewrite Equation~\ref{eq:bulk_ode} as:
\begin{equation}
    \frac{1}{2}\left(\sigma_0u' + c\right)^2 = \frac{c^2}{2} - \left(1 -\frac{z^2}{2} - \rho_K\right) = \frac{c^2}{2} - R(z), \label{eq:completed_square}
\end{equation}
where $R(z)$ is the fitness of trait $z$.

As the right-hand side will always be nonnegative, we can deduce that $\rho_K\geq1-\frac{c^2}{2}-\frac{z^2}{2}$. In case of equality at the tip, we have what is called a pulled front, as the F-KPP relationship $\frac{c^2}{2}=R(z_K)=1-\frac{z_K^2}{2} - \rho_K$ (\cite{roques2012}, see also~\cite{murray2002} for an introductory discussion of the F-KPP equation) is satisfied at the right-most trait. In a pulled front, dynamics are driven by the tip of the population. Further, in this case we have the sharpest decay of the profile near the tip, given by $u'(z_K)=-\frac{c}{\sigma_0}$. In the case of a strict inequality, we have a pushed front, characterised by $\frac{c^2}{2}>R(z_K)=1-\frac{z_K^2}{2} - \rho_K$. Such a front will be driven by the bulk of the population. In this cased we would also see milder decay at the tip, with $u'(z_K) < -\frac{c}{\sigma_0}$. We make the assumption that the front is pulled when $z_K<0$, and as we will see in Section~\ref{sec:trajectories}, lineage arguments predict that the population will descend from individuals at the tip. Further, this corresponds to the highest possible rate of growth when $z_K<0$. Thus, we have:
\begin{equation*}
    \rho_K^0 = 1 - \frac{c^2}{2}-\frac{z_K^2}{2},
\end{equation*}
which is our leading-order prediction of the population size based on the bulk. This prediction corresponds to the bulk prediction we will derive in Section~\ref{sec:trajectories},

Returning to Equation~\ref{eq:completed_square}, we further see that:
\begin{equation*}
    u'(z) = -\frac{c}{\sigma_0} \pm \frac{1}{\sigma_0} \sqrt{\rho_K - 1 +\frac{c^2}{2}+\frac{z^2}{2}}.
\end{equation*}
By the expected continuity of $u'(z)$ due to the regularity of viscosity solutions, we must choose to either add or subtract the square root term. To ensure $u(z)$ reaches a maximum, that is, that $u''(z)<0$, we must choose the addition of the square root. Therefore:
\begin{equation*}
    u'(z) = -\frac{c}{\sigma_0} + \frac{1}{\sigma_0} \sqrt{\rho_K - 1 +\frac{c^2}{2}+\frac{z^2}{2}} \geq -\frac{c}{\sigma_0},
\end{equation*}
and further the solution for $u(z)$ is given by:
\begin{equation}
    u(z)=\underbrace{\frac{1}{2\sigma_0}\left((2\rho_K+c^2-2)\ln|\sqrt{c^2+z^2+2\rho_K-2} + z| + z(\sqrt{c^2+z^2+2\rho_K-2} - 2c) \right)}_{v(z)} + C. \label{eq:bulk_solution}
\end{equation}
Applying the initial condition $u(z_K)=0$ and plugging in our prediction for $\rho_K^0$, the complete solution becomes:
\begin{equation}
    u(z) = \frac{1}{\sigma_0}\left[c(z_K-z) + \frac{z\sqrt{z^2-z_K^2}}{2} - \frac{z_K\ln\left|\frac{\sqrt{z^2-z_K^2}+z}{z_K}\right|}{2}\right]. \label{eq:leading_order_bulk_solution}
\end{equation}

To find the value of $z_K$, we impose the self-consistency condition that $\int_{-\infty}^{z_K}K^{u(z)-1}dz=\rho_K^0$. This can be approximated with the Laplace method yielding the condition:
\begin{equation*}
    \sqrt{\frac{2\pi\sigma_0c}{\ln (K)|u''(z_{max}|}}\exp\left(\ln (K) (u(z_{max}-1)\right)=\rho_K^0,
\end{equation*}
where $z_{max}$ is such that $u'(z_{max})=0$. This can be solved numerically for $z_K$ for all values of $c>c_{trans}$. However, as our conditions on the equation have changed, the transition speed derived in Section~\ref{sec:supp_ctrans} will not be compatible with our analysis. To define a transition speed that is consistent with our conditions, we first take the limit of Equation~\ref{eq:leading_order_bulk_solution} as $z_K\to 0$. This yields:
\begin{equation*}
    \frac{1}{2}\ln(2\pi\sigma_0c)-\frac{1}{2}\ln(\ln K)-\frac{1}{2}\ln c + \frac{c^2}{2\sigma_0}\ln K-\ln K = \ln\left(1-\frac{c^2}{2}\right),
\end{equation*}
which can be solved numerically for a leading order $c_{trans}$. Hence for $c\leq c_{trans}$ we define $\rho_K^0=1-\frac{c^2}{2}$ and for $c>c_{trans}$, we first solve numerically for $z_K$, then set $\rho_K^0=1-\frac{c^2}{2}-\frac{z_K^2}{2}$. Additionally, we plot the prediction of $\rho_K^0$ when simulation values of the right-most trait $Z_K$ are substituted in for $z_K$ in $\rho_K^0=1-\frac{c^2}{2}$, as detailed in Section~\ref{sec:predicting_rho}. Figure~\ref{fig:leading_order} shows both of these predictions compared with the uncorrected deterministic prediction $\rho_{\infty}$ and observed population sizes in simulations. We see that while our equation for the population size performs relatively well if $z_K$ is known, this analysis incorrectly predicts the transition speed and the location of $z_K$.

To improve this prediction, we might first look at the next order correction of our equation on $u(z)$ involving $\varepsilon\sigma_0^2u''(z)$. However, we see that the second derivative of $u(z)$ is not defined at $z_K$. This motivates us to seek a correction by instead considering a boundary layer when $z\lesssim z_K$. In particular, we seek to define a solution at the tip near $z_K$ and choose $C$ such that Equation~\ref{eq:bulk_solution} matches the tip solution at some match point $z_{match}$. We can then use the combination of these two solutions to make a new prediction of $z_K$ and $\rho_K$ for values of $c>c_{trans}$, where $c_{trans}$ is once again defined as the value derived in Section~\ref{sec:supp_ctrans}.

\subsubsection{Solution to the linearised problem at the tip} \label{sec:supp_linearised}

Consider Equation~\ref{eq:ogODE} at the tip when $z\lesssim z_K$. We first expand the non-linear term near $z_K$ to get:
\begin{equation*}
    \sigma^2f''(z) + 2\sigma c f'(z) + (2 - 2\rho_K - z_K^2 - 2z_K(z-z_K) - (z-z_K)^2)f(z)=0.
\end{equation*}
Letting $x := z-z_K$, this can be rewritten as:
\begin{equation}
    \sigma^2f''(x) + 2\sigma c f'(x) + (2 - 2\rho_K - z_K^2 - 2z_Kx - x^2)f(x)=0. \label{eq:fx_big}
\end{equation}
Since $z\lesssim z_K$, this implies that the variable $x$ must be small. Suppose that $x = \mathcal{O}(\varepsilon^\alpha)$, for some exponent $\alpha$ to be determined. Then $x$ can be written as $x:=\varepsilon^\alpha y$, where $y=\mathcal{O}(1)$. The goal now becomes to choose $\alpha$ such that the non-linear term of our equation is the only term of order less than 1. Substituting in $y$, we arrive at:
\begin{equation*}
    \sigma^2\varepsilon^{-2\alpha}f''(y) + 2\sigma c \varepsilon^{-\alpha} f'(y) + (2 - 2\rho_K - z_K^2 - 2z_K\varepsilon^\alpha y - \varepsilon^{2\alpha}y^2)f(y)=0.
\end{equation*}
Setting $\sigma = \sigma_0\varepsilon$ and multiplying through by $\varepsilon^{2\alpha-2}$ this becomes:
\begin{equation*}
    \sigma_0^2f''(y) + 2\sigma_0 c \varepsilon^{\alpha-1} f'(y) + (\varepsilon^{2\alpha -2}(2 - 2\rho_K - z_K^2) - 2z_K\varepsilon^{3\alpha-2} y - \varepsilon^{4\alpha-2}y^2)f(y)=0.
\end{equation*}
We see here that in choosing $\alpha = 2/3$, the non-linear term $\varepsilon^{4\alpha-2}y^2$ will be the only term of order less than one. Hence our linearised ODE is:
\begin{equation*}
    \sigma_0^2f''(y) + 2\sigma_0 c \varepsilon^{-1/3} f'(y) + (\varepsilon^{-2/3}(2 - 2\rho_K - z_K^2) - 2z_K y)f(y) = 0. 
\end{equation*}
The scaling of the boundary layer is therefore of size $\varepsilon^{-2/3}$, therefore any matching point in $x$ must be of this order. Letting $f(y)=\exp\left(-\frac{c}{\sigma_0}\varepsilon^{-1/3}y\right)g(y)$, we can factor out this exponential term from the equation to arrive at:
\begin{equation*}
    \left(\sigma_0^2g''(y) + (\varepsilon^{-2/3}(2-2\rho_K-z_K^2+c^2)-2z_Ky)g(y)\right)\exp\left(-\frac{c}{\sigma_0}\varepsilon^{-1/3}y\right)=0.
\end{equation*}
We note two things. First that the exponential term is identical to the first term in Equation~\ref{eq:leading_order_bulk_solution} when we return to the variable $z$ by the change of variables $y=\varepsilon^{-2/2}(z-z_K)$. Second, that the equation in $g(y)$ is the Airy equation, subject to a change of variables. In particular, let $\alpha = \frac{\gamma'-y}{\delta'}$. Then the equation in $g(y)$ becomes $g''(\alpha)-\alpha g(\alpha)=0$ when $\gamma'=\frac{\varepsilon^{-2/3}(c^2-2+2\rho_K+z_K^2)}{|2z_K|}$ and $\delta'=\left(\frac{\sigma_0^2}{|2z_K|}\right)^{1/3}$. We choose the first Airy solution $g(\alpha)=A\mathrm{Ai}(\alpha)$ so that our solution is biologically valid in the case of linear mortality (that is, $f(z)\to 0$ as $z\to-\infty$, which requires that $g(\alpha)\to0$ as $\alpha\to\infty$). Further, this choice of the Airy solution best matches the behaviour of the bulk solution asymptotically. Therefore, $f(y)=A\exp\left(-\frac{c}{\sigma_0}\varepsilon^{-1/3}y\right)\mathrm{Ai}\left(\frac{\gamma'-y}{\delta'}\right)$. Returning to the variable $x$ by the change of variables $y=\varepsilon^{-2/3}x$, the solution is therefore:
\begin{equation*}
    f(x) = A\exp\left(-\frac{c}{\sigma_0\varepsilon}x\right)\mathrm{Ai}\left(\frac{\gamma - x}{\delta} \right ),
\end{equation*}
where:
\begin{equation*}
    \gamma = \frac{c^2-2+2\rho_K+z_K^2}{|2z_K|},
\end{equation*}
and:
\begin{equation*}
    \delta = \left(\frac{\sigma_0^2\varepsilon^2}{|2z_K|}\right)^{1/3}.
\end{equation*}
Note that this is the solution to Equation~\ref{eq:fx_big} when the quadratic term is dropped~\citep{cohen2005,roques2017}. Since we impose that the population density must reach 0 at $z_K$, this implies:
\begin{equation*}
    f(0) = A\exp (0) \mathrm{Ai}\left(\frac{\gamma}{\delta}\right) = 0.
\end{equation*}
This implies that $\frac{\gamma}{\delta} = \xi_0\approx -2.34$, the first zero of the Airy function. We choose the first zero to maintain a positive population density. This provides an updated formula for $\rho_K$, as $\frac{\gamma}{\delta} = \xi_0$ implies:
\begin{equation*}
    \frac{c^2-2+2\rho_K+z_K^2}{|2z_K|} = \varepsilon^{2/3}\xi_0\left(\frac{\sigma_0^2}{|2z_K|}\right)^{1/3},
\end{equation*}
from which we can conclude that $\rho_K$ is given by Equation~\ref{eq:lambdaK}. Note that without imposing any additional assumptions, we have recovered the leading order equation $\rho_K^0 = 1-\frac{c^2}{2}-\frac{z_K^2}{2}$ which we proposed in Section~\ref{sec:supp_bulk}. Figures~\ref{fig:range}B and~\ref{fig:big_var_range} show that substituting in simulation values $Z_K$ for $z_K$, Equation~\ref{eq:lambdaK} predicts observed $\frac{N}{K}$ well, with the addition of the smaller-order term improving the prediction (compare with Fig.~\ref{fig:leading_order}).

To find the distribution at the tip, it remains to find the constant $A$. 
As when solving for $c_{trans}$, we make the heuristic argument that the density at the very tip of the population distribution between $z_K-\bar{s}$ and $z_K$ should be $1/K$, corresponding to one individual:
\begin{equation*}
    K\int_{-\bar{s}}^0f(x)dx=K\int_{-\bar{s}}^0A\exp\left(-\frac{c}{\sigma}x\right)\mathrm{Ai}\left(\xi_0 - \left(\frac{|2z_K|}{\sigma^2}\right)^{1/3}x\right)dx = 1,
\end{equation*}
(see again Sec.~\ref{sec:sbar} for calculation and interpretation of $\bar{s}$). Approximating $\mathrm{Ai}\left(\xi_0 - x\left(\frac{|2z_K|}{\sigma^2}\right)^{\frac{1}{3}}\right)$ with $-\left(\frac{|2z_K|}{\sigma^2}\right)^{\frac{1}{3}}\mathrm{Ai}'(\xi_0)x$, we get:
\begin{align*}
    \frac{1}{K} &\approx -A\left(\frac{|2z_K|}{\sigma^2}\right)^{\frac{1}{3}}\mathrm{Ai}'(\xi_0)\int_{-\bar{s}}^0x\exp\left(-\frac{c}{\sigma}x\right)du \\
    &= A \left(\frac{|2z_K|}{\sigma^2}\right)^{\frac{1}{3}}\mathrm{Ai}'(\xi_0)\frac{\left(\bar{s}\frac{c}{\sigma}-1\right)\exp\left(\bar{s}\frac{c}{\sigma}\right)+1}{\frac{c^2}{\sigma^2}}
\end{align*}
Therefore, we find the constant $A$ is given by:
\begin{equation}
    A = \frac{c^2}{\sigma^{4/3}K\mathrm{Ai}'(\xi_0)|2z_K|^{1/3}\left(\left(\frac{\bar{s}}{\sigma}c-1\right)\exp\left(\frac{\bar{s}}{\sigma}c\right)+1\right)},
    \label{eq:A_fk}
\end{equation}
and so our solution at the tip is:
\begin{equation}
    f_T(x)= \frac{c^2}{\sigma^{2/3}K\mathrm{Ai}'(\xi_0)|2z_K|^{1/3}\left(\left(\frac{\bar{s}}{\sigma}c-1\right)\exp\left(\frac{\bar{s}}{\sigma}c\right)+1\right)}\exp\left(-\frac{c}{\sigma}x\right)\mathrm{Ai}\left(\xi_0 - \left(\frac{|2z_K|}{\sigma^2}\right)^{1/3}x\right). \label{eq:tip_solution}
\end{equation}

\subsubsection{Matching the bulk to the tip} \label{sec:supp_match}

To match our two solutions, we first require a matching point. We note that the solution on the bulk $u(z)$ is only defined for $z\leq -\sqrt{2-c^2-2\rho_K}$, so we set this as our match point. We therefore wish to find the integration constant $C$ from Equation~\ref{eq:bulk_solution} such that Equation~\ref{eq:bulk_solution} and Equation~\ref{eq:tip_solution} agree at $z_{match}=-\sqrt{2-c^2-2\rho_K}$. That is:
\begin{equation*}
    K^{u(z_{match})-1}=f_T(x_{match}),
\end{equation*}
where:
\begin{align}
    x_{match} &= z_{match}-z_K \nonumber \\
    &= -z_K-\sqrt{2-c^2-2\rho_K} \nonumber \\
    &= -z_K-\sqrt{z_K^2-\varepsilon^{2/3}\xi_0(2\sigma_0z_K)^{2/3}} \nonumber \\
    &\approx \varepsilon^{2/3}\xi_0\frac{\sigma_0^{2/3}}{|2z_K|^{1/3}}. \label{eq:match_point}
\end{align}
We note that the scale of our matching point $x_{match}$ is the same as the scale of our tip solution, which is defined on $x=\mathcal{O}(\varepsilon^{2/3})$. In other words, our matching point is sufficiently close to $z_K$ for both solutions to hold at this point. Therefore, we can calculate $C$ as:
\begin{align*}
    C &= \varepsilon\ln\left[f_T\left(x_{match}\right)\right]+1-v(z_{match}),
\end{align*}
where $v(z)$ is the known piece of $u(z)$ (Eq.~\ref{eq:bulk_solution}). Since this computation is done numerically, there is no need for approximation and we use the exact value of $x_{match}$ for precision.

\subsubsection{Finding $z_K$} \label{sec:supp_find_zk}

Finally, to determine $z_K$, we choose it to be such that:
\begin{equation*}
    \rho_K = \int_{-\infty}^{z_K} f(z)dz \approx \int_{-\infty}^{z_{match}} e^{\ln K(u(z)-1)}dz,
\end{equation*}
which gives us the value of $z_K$ such that the density of the population in the bulk is equal to the predicted population density $\rho_K$. To solve this equation, we approximate the integration over the bulk with Laplace's method centered around the maximum of the distribution, $z_{max} = -\sqrt{2-2\rho_K}$ (given by $u'(z_{max})=0$, see Eq.~\ref{eq:bulk_ode}):
\begin{equation*}
    \int_{-\infty}^{z_{match}} e^{\ln K(u(z)-1)}dz \approx \sqrt{\frac{2\pi}{\ln K |u''(z_{max})|}}\exp\left(\ln K(u(z_{max})-1\right).
\end{equation*}
We then set this equation equal to $\rho_K$ as given by Equation~\ref{eq:lambdaK}, giving us:
\begin{equation}
    \sqrt{\frac{2\pi}{\ln K |u''(z_{max})|}}\exp\left(\ln K(u(z_{max})-1\right) = 1-\frac{c^2}{2} - \frac{z_K^2}{2} + \varepsilon^{2/3}\frac{\xi_0}{2}|2z_K\sigma_0|^{2/3}.\label{eq:root_eq}
\end{equation}
This must be solved numerically for $z_K$ using a root-finding algorithm. 

It must be noted that the derivation of this equation assumes $z_K$ to be sufficiently far from zero. This corresponds to speeds which are sufficiently greater than the transition speed $c_{trans}$, with $c_{trans}$ as calculated in Section~\ref{sec:supp_ctrans}. As such, we only calculate the values of $z_K$ for speeds sufficiently greater than $c_{trans}$.
For a given value of $c$, we first calculate $u(z_{max})$ and $u''(z_{max})$ from Equation~\ref{eq:bulk_solution}. Using these values, we solve for $z_K$ as the root of Equation~\ref{eq:root_eq}. Finally, we plug this value into Equation~\ref{eq:lambdaK} to find $\rho_K$. In other words, for a given $c$, we must find $z_K$ such that the following equations hold:
\begin{equation*}
    \begin{cases}
u(z)=\frac{1}{2\sigma_0}\left((2\rho_K+c^2-2)\ln|\sqrt{c^2+z^2+2\rho_K-2} + z| + z(\sqrt{c^2+z^2+2\rho_K-2} - 2c) \right) + C \\
f_T(z)= \frac{c^2}{\sigma^{2/3}K\mathrm{Ai}'(\xi_0)|2z_K|^{1/3}\left(\left(\frac{\bar{s}}{\sigma}c-1\right)\exp\left(\frac{\bar{s}}{\sigma}c\right)+1\right)}\exp\left(-\frac{c}{\sigma}(z-z_K)\right)\mathrm{Ai}\left(\xi_0 - \left(\frac{|2z_K|}{\sigma^2}\right)^{1/3}(z-z_K)\right) \\
\rho_K = 1- \frac{\tilde{c}^2}{2\sigma^2} - \frac{z_K^2}{2} + \frac{\xi_0}{2}|2z_K\sigma|^{2/3} \\
z_{match}=-\sqrt{2-c^2-2\rho_K} \\
K^{u(z_{match})-1}=f_T(z_{match})\\
u'(z_{max}) = 0\\
\sqrt{\frac{2\pi}{\ln K |u''(z_{max})|}}\exp\left(\ln K(u(z_{max})-1\right) = \rho_K
\end{cases}
\end{equation*}
Note that due to the approximations made to maintain tractability, the root of Equation~\ref{eq:root_eq} does not exist for large enough values of $c$ (see Fig.~\ref{fig:supp_range}).

\section{Simulation framework} \label{sec:simulations}

As we want to consider simulations with large numbers of individuals, implementing a Gillespie algorithm for our individual-based model was impractical. Instead, we consider a discrete-time simulation where every individual is updated at every time step of length $dt$. We take our simulation to be in the moving frame. In each time step, the phenotypic values of every individual must drift from the optimal trait $0$ by $\tilde{c}dt = \sigma c dt$. Individuals give birth at rate $1$, and die at rate $\frac{z^2}{2}+\frac{N}{K}$. Over the course of a time step, these birth and death rates correspond to the probability of an individual having a single offspring, $1-e^{-dt}$, and the probability of an individual surviving to the next time step, $\exp\left(-\left(\frac{z^2}{2}+\frac{N}{K}\right)dt\right)$, where $N$ is the total number of individuals. In the event of a birth, the offspring will mutate with probability $\eta$. The size of this mutation is drawn from a normal distribution with mean 0 and variance $\sigma^2_{mut}$. Note that $\eta$ and $\sigma^2_{mut}$ are such that $\sigma^2=\eta\sigma^2_{mut}$. 

A simulation proceeds as follows. The size of time step is set as $dt=\frac{1}{1000\sigma}$. The simulation is initialised with a list of $K$ individuals drawn randomly from a normal distribution with mean 0 and variance $\sigma^2$. With each time step, the trait value of every individual is first shifted back by $\frac{1}{2}\sigma cdt$. Each individual then reproduces with probability $1-e^{-dt}$, independently of each other. Each offspring undergoes a mutation with probability $\eta$. For offspring which mutates, the mutation is drawn from a distribution of mutational effect sizes with mean 0 and variance $\sigma_{mut}^2$. Each parent then survives with probability $\exp\left(-\left(\frac{z^2}{2}+\frac{N}{K}\right)dt\right)$. The updated list of individuals containing surviving parents and new offspring is then shifted back by $\frac{1}{2}\sigma cdt$.

In asexual populations, mutations are often modelled analytically as diffusion~\citep{kimura1965}. Within a simulation framework, this means that the offspring does not undergo a single large mutation at birth. Instead, at every time step a random mutation is drawn from a normal distribution with mean 0 and variance $\sigma^2 dt$ for each individual in the updated list of containing surviving parents and new offspring. Figure~\ref{fig:mut_rates} shows that for frequent mutations ($\eta\simeq1$) and small $\sigma^2_{mut}$, simulations with normally distributed mutations resemble those modelled with diffusion. Thus, throughout our work we assume the diffusion approximation in the PDE and use the more biologically reasonable simulations where mutations are drawn from a Gaussian distribution at probability $\eta=1$ and with variance $\sigma_{mut}^2=\sigma^2$.

In sexual populations, for every birth event, a second parent is chosen a random. Segregation and recombination can then be approximated with an infinitesimal operator, which assumes a large number of additive loci with limited linkage~\citep{barton2017}. In this case, the trait value of the offspring is then the mean value of its parents plus some random variation drawn from a Gaussian distribution with mean 0 and variance $\frac{\sigma^2}{2}$. Additionally, in the sexual case time is scaled by $\sigma^2$ as opposed to $\sigma$~\citep{garnier2023}. As such, the time step is given by $\frac{1}{1000\sigma^2}$. The rate of environmental change is scaled as $\tilde{c}=\sigma^2c$ and so the trait value of every individual is shifted back by $\frac{1}{2}\sigma^2cdt$.

Figure~\ref{fig:ace_v_allo} takes $\sigma_{asex}=0.01$ for the asexual population and $\sigma_{sex}=0.1$ for the sexual population so that simulations for both populations could be performed on the same time scale with the same absolute rates of environmental change while capturing the transition in population size. In the asexual population, mutations are drawn from a normal distribution. Additionally, a carrying capacity of $K=10000$ was used. Simulations were run for 15~000 time steps and 15 replicates. Figure~\ref{fig:supp_ace_v_allo} shows that the discrepancy between asexual and sexual populations holds across parameter values. 

\subsection{Predicting $\rho_K$ from $Z_K$} \label{sec:predicting_rho}

In order to show how well our predictions perform for $\rho_K$ (Eq.~\ref{eq:lambdaK} and its leading order, $1-\frac{c^2}{2}-\frac{z_K^2}{2}$), we want to plot these predictions given simulation measurements of $Z_K$. At the end of each simulation run, the right-most trait ($Z_K$) was recorded. Then across each replicate, if the right-most trait was negative, we calculated the predicted $\rho_K$ when taking $z_K$ to be this measured trait $Z_K$. If the right-most trait was positive, $\rho_K$ was taken as the infinite-population prediction. If the population had gone extinct, we set $\rho_K$ to be zero. Figures~\ref{fig:range}B,~\ref{fig:leading_order}, and~\ref{fig:big_var_range} show the mean and standard deviation of all these $\rho_K$ values.

\subsection{Mutational effect sizes at the tip}

In \cite{roques2017} it is assumed that the step forward of the distribution through the trait space corresponds with the expected size of mutation from the tip, $\bar{s}=\sigma$. We see in Figure~\ref{fig:range}B that this assumption does not yield an accurate prediction for the finite-population dynamics in our model. To understand this discrepancy and test our alternative calculations of $\bar{s}$ (Sec.~\ref{sec:sbar}), it is necessary to investigate the size of $\bar{s}$ and its relationship to the progression of the phenotypic distribution in simulations.

To better understand the stochastic behaviours at the tip of the trait distribution and the nature of the mutations that drive evolution, we first run simulation for 20~000 time steps to arrive at a quasi-stationary state, then run for a further 10~000 time steps while sampling the state of the tip after each time step. With each instance, we track the size of the mutation between the new right-most trait and its parent. This measurement does not take into account whether the new right-most trait establishes, however the probability of establishment is relatively flat over the range in which mutations arise, and thus does not weight things significantly. Figure~\ref{fig:step_sizes} shows the average mutational effect size as well as the average distance between the new right-most trait and the previous right-most trait (in the fixed frame). We see that the size of mutations is much larger than the distance between the new right-most trait and its predecessor, indicating that on average the parent that gives rise to the new right-most trait is found behind the right-most trait. The average step forward in the distribution is smaller than $\sigma$, while the average effect size is on the order of $\sigma$, but not as large as $2\sigma$ for all parameter values we looked at. It is also interesting to note that the effect size does not show a significant change between small speeds–where the optimal trait is still found in the population–and high speeds–where the optimal trait has been lost. However, this might change if these measurements were limited to tracking the right-most individuals that establish, as right-most traits are less likely to establish if they are located past the optimal trait as opposed to when they are approaching the optimal trait from behind.

\section{Determining $\bar{s}$} \label{sec:sbar}

Our analysis thus far has depended on the value $\bar{s}$, which is meant to characterise the expected size of the step forward through the trait space taken by the population when a mutation creates a new individual at the right-most tip of the distribution. Drawing from this behaviour in individual-based simulations, we follow the heuristic argument made by~\cite{roques2017} that the density over this interval at the tip should correspond to one individual. That is, constants should be chosen such that:
\begin{equation*}
    K\int_{z_K-\bar{s}}^{z_K}f(z)dz = 1.
\end{equation*}
As in~\cite{roques2017}, we start with the assumption that $\bar{s}=\sigma$. In this case, the size of the step forward is the average effect of mutations from the parent. The new right-most trait is always assumed to be the offspring of a parent carrying the right-most trait. As Figure~\ref{fig:range}B shows, this yields a correction which approaches the simulation results, but overestimates the fitness loss due to finite population effects (see Fig.~\ref{fig:supp_range}, $L=1$). However, by considering a range of values of the order of $\sigma$, we find that the simulation results can be well-approximated by choosing some $\bar{s}=L\sigma$. The best match appears to be around $\bar{s}\approx1.7\sigma$. This shows some dependence of parameters $c$, $\sigma$, and $K$, but remains relatively close to this value.

When an offspring takes the place of the new right-most trait, the average size of the mutational step from the parent to the offspring (Fig.~\ref{fig:step_sizes}, purple dots) is greater than $\sigma$. It is also greater than the size of the step forward of the distribution of traits through the trait space in the fixed frame (Fig.~\ref{fig:step_sizes}, green crosses). 
In order to arrive at a more accurate prediction for $\rho_K$, we consider that mutations which progress the trait distribution may arise from parental traits behind the tip, $z_K$. Therefore, we consider $\bar{s}$ to instead be the expected effect size of mutations that establish, conditioned on the offspring carrying a trait closer to the optimum than $z_K$. 

\subsection{Expected effect size of mutations that fix under rapid mutation} \label{sec:first_sbar}

For simplicity, we first assume the mutations are arising rapidly such that there is no waiting time until the next mutation that establishes. In this case, $\bar{s}$ should be given by:
\begin{equation}
    \bar{s} = \frac{\int_{-\infty}^{z_K}\int_{|z-z_K|}^\infty s f(z)\lambda(s)w(z+s)dsdz}{\int_{-\infty}^{z_K}\int_{|z-z_K|}^\infty f(z)\lambda(s)w(z+s)dsdz},  \label{eq:simp_sbar}
\end{equation}
where $\lambda(s)$ is the probability distribution of mutational effect sizes and $w(z)$ is the establishment probability of a lineage starting with trait $z$. The size $s$ of the effect size must be large enough to reach the tip $z_K$, hence at least of size $|z-z_K|$.

To find $\bar{s}$ in this case, we must first look for the establishment probability of an individual with trait $z$. We start with the most simple case, which is to assume that every individual establishes. That is, $w(z)=1$ and $\lambda(s)$ a normal distribution with mean 0 and variance $\sigma^2$. 

For a given speed $c$, we find little variation in the value of $\bar{s}$ across a range of values of $z_K$ (Fig.~\ref{fig:sbar_over_zk}). As such, when solving for $z_K$ in Equation~\ref{eq:root_eq}, we do not substitute in $\bar{s}$ as a general function of $z_K$. For computational efficiency, we instead calculate $\bar{s}$ for a range of possible $z_K$ values and substitute the mean of these values into Equation~\ref{eq:root_eq}. This measure of $\bar{s}$ defined by Equation~\ref{eq:simp_sbar} yields a significantly improved prediction from setting $\bar{s}=\sigma$, shown in Figure~\ref{fig:w1correction}. In Section~\ref{sec:find_w} we consider a more complex case when the establishment probability is no longer 1, but find that this does not significantly change the prediction.

While this calculation of $\bar{s}$ consistently provides a more accurate match with simulation results than $\bar{s}=\sigma$, its interpretation is unclear. We see in Figure~\ref{fig:step_sizes} that this prediction of the expected effect size does not match the observed effect sizes in simulation, which are smaller than $\bar{s}$, the values of which are shown in Figure~\ref{fig:sbar_over_zk}. Additionally, it is unclear what it means to integrate over $z$ on this interval. By considering mutations that come from behind the tip, the expected effect size is no longer equal to the step forward taking by the trait distribution. This means that this interval should span a density of more than one individual, since the offspring is ``leapfrogging'' over individuals that exist between its parent and the right-most individual. These limitations, combined with the surprising success of the predictions given by $\bar{s}$, suggest that although this is not the correct approach to refining our prediction, it does capture some signal for what the correct approach may be.

\subsection{Calculating the establishment probability of a new mutation} \label{sec:find_w}

In this section, we refine the calculation presented in Section~\ref{sec:first_sbar} by incorporating the fact that not all offspring that become the new right-most individual will successfully establish.
Following the methodology of~\cite{good2012}, we define the establishment probability at time $t$ as:
\begin{equation*}
    w(x,t) = 1 - p(1,x,t),
\end{equation*}
where $p(n,x,t)$ is the extinction probability of a lineage with $n$ individuals carrying trait $x$ at time $t$. This probability will follow the backward master equation:
\begin{multline*}
    p(n,x,t-dt) = \underbrace{\left [1 - ndt \left(1 + \frac{(x-\sigma ct)^2}{2} + \rho_K \right ) \right]}_{nothing}p(n,x,t) + \underbrace{ndt \left(\frac{(x-\sigma ct)^2}{2} + \rho_K \right)}_{death}p(n-1,x,t) \\ + \underbrace{ndt\int\lambda(s)p(1,x+s,t)ds}_{birth\ w/ \ mutation}p(n,x,t),
\end{multline*}
where $\lambda(s)$ is the probability distribution of effect sizes. In the continuous-time limit this becomes:
\begin{multline*}
    -\frac{1}{n}\partial_tp(n,x,t)=-\left(1 + \frac{(x-\sigma ct)^2}{2} + \rho_K \right )p(n,x,t) + \left(\frac{(x-\sigma ct)^2}{2} + \rho_K \right)p(n-1,x,t) \\ + \int\lambda(s)p(1,x+s,t)ds p(n,x,t). \label{eq:n_master}
\end{multline*}
Since $n=0$ or $n=1$ (as every mutation creates a new lineage), and $p(0,x,t)$ is clearly 1, for $n=1$ this becomes:
\begin{multline*}
    -\partial_tp(1,x,t) = -\left(1 + \frac{(x-\sigma ct)^2}{2} + \rho_K \right )p(1,x,t) + \left (\frac{(x-\sigma c t)^2}{2} + \rho_K \right ) \\ + \int\lambda(s)p(1,x+s,t)ds p(1,x,t).
\end{multline*}
Substituting in $w(x,t)$, we arrive at:
\begin{equation*}
    \partial_tw = \left(\frac{(x-\sigma ct)^2}{2} + \rho_K \right )w - (1-w)\int \lambda(s)w(x+s,t)ds.
\end{equation*}
In the moving frame $z=x-\sigma ct$, this becomes the ODE:
\begin{equation*}
    -\sigma cw'=\left(\frac{z^2}{2} + \rho_K\right)w-(1-w)\int \lambda(s)w(z+s)ds.
\end{equation*}
We take the diffusion approximation to arrive at:
\begin{equation}
    \underbrace{-\sigma cw'}_{O} = \underbrace{\left(1 - \frac{z^2}{2} - \rho_K \right )w}_{S} + \underbrace{(1-w)\frac{\sigma^2}{2}w''}_M - \underbrace{w^2}_N.
    \label{eq:wODE}
\end{equation}
Here we have labelled the moving optimum term (O), the selection term (S), the mutation term (M), and the nonlinear term (N).

Equation~\ref{eq:wODE} cannot be solved explicitly, so we instead consider multiple scales. For traits far from the optimal trait, the establishment probability will tend to zero. Near the optimal trait, where fitnesses are high, lineages will generally establish or die before mutation or a change in the optimum can alter their fitness dramatically. This assumes In this case the selection (S) and nonlinear (N) terms will dominate, yielding the approximate solution:
\begin{equation}
    w_I(z)\approx 1 - \frac{z^2}{2} - \rho_K,
    \label{eq:inst_fix}
\end{equation}
which is the instantaneous establishment probability (which we denote by subscript $I$). 

Lagging behind the optimal trait, for values near $z_K$, we can linearise. Applying the same substitution $y=\varepsilon^{2/3}(z-z_K)$ as in the case of $f(z)$ near $z_K$ and letting $\sigma = \varepsilon\sigma_0$, we have:
\begin{equation*}
    \varepsilon^{1/3}\sigma_0cw'=(1-\varepsilon^{4/3}\frac{y^2}{2} - \varepsilon^{2/3}z_Ky-\frac{z_K^2}{2}-\rho_K)w + \varepsilon^{2/3}(1-w)\frac{\sigma_0^2}{2}w''-w^2.
\end{equation*}
Multiplying through by $\varepsilon^{-2/3}$, the nonlinear term is the only term of order less than one, so we drop this term. Substituting back $x=\varepsilon^{2/3}y$ and multiplying through by $\varepsilon^{2/3}$, we get:
\begin{equation*}
    \varepsilon\sigma_0cw'=(1-z_Kx-\frac{z_K^2}{2}-\rho_K)w + \varepsilon^2(1-w)\frac{\sigma_0^2}{2}w'' - w^2.
\end{equation*}
Here, the mutation term is of order $\mathcal{O}(\varepsilon^2)$. Behind the optimal trait, we also expect the establishment probability and thus the nonlinear term to be negligible. Hence the moving optimum term (O) and the selection term (S) dominate. In this case, we arrive at the approximate solution:
\begin{equation}
    w_L(x) \approx A\exp\left(-\frac{z_Kx^2-z_K^2x-2(1-\rho_K)x}{2\sigma c} \right),
    \label{eq:exp_fix}
\end{equation}
which we call the long-term establishment probability, denoted by the subscript $L$.

Around the transition speed, when $z_K=0$, we expect the success of new mutations past $z_K$ to be determined by the instantaneous establishment probability (Eq.~\ref{eq:inst_fix}). Hence the expected effect size of mutations that establish, conditioned on the offspring's trait being greater than $z_K$ is given by:
\begin{equation*}
    \bar{s} = \frac{\int_{-\infty}^{0}\int_{-z}^\infty s f_0(z)\lambda(s)w_I(z+s)dsdz}{\int_{-\infty}^{0}\int_{-z}^\infty f_0(z)\lambda(s)w_I(z+s)dsdz},
\end{equation*}
where $\lambda(s)$ is a Gaussian distribution of effect sizes with mean 0 and variance $\sigma^2$.

Past the transition speed, when $z_K$ is sufficiently far from zero, we expect the success of new mutations past $z_K$ to instead be determined by the long-term establishment probability (Eq.~\ref{eq:exp_fix}). In this case, the expected effect size of mutations that establish past $z_K$ is given by:
\begin{equation*}
    \bar{s} = \frac{\int_{x_{match}}^{0}\int_{-x}^\infty s f_T(x)\lambda(s)w_L(x+s)dsdx}{\int_{x_{match}}^{0}\int_{-x}^\infty f_T(x)\lambda(s)w_L(x+s)dsdx},
\end{equation*}
where we recall $x = z-z_K$. Here we only integrate over the tip solution, as we expect the contribution of large-effect mutations from parental traits in the bulk to be negligible. Setting the lower limit on the inner integral as $-x=z_K-z$ constrains us to positive mutation effect sizes that are large enough for the offspring to have a trait value greater than $z_K$, hence these are mutations that will drive the population distribution forward through the trait space. This measure of $\bar{s}$ yields the prediction shown in Figure~\ref{fig:wcorrection}. The accuracy of this prediction is limited by the approximations made to derive $w(z)$. The prediction for the instantaneous establishment probability $w_I$ may not be accurate if $c$ is too close to 0 and thus $\rho_K$ too large, in which case the establishment probability may be small. However, we note the similarities to our predictions in Figure~\ref{fig:w1correction}, suggesting the establishment probability does not play an important role in determining $\bar{s}$.

\subsection{Expected effect size of mutations that after some waiting time}

Our next attempt at defining $\bar{s}$ still considers the expected effect size of mutations coming from anywhere in the distribution, 
For tractability, our calculation of the expected effect size of mutations coming from anywhere in the distribution as presented in Sections~\ref{sec:first_sbar} and~\ref{sec:find_w} assumes rapid mutations. However, by ignoring waiting times between mutations, we avoid the environmental and consequent fitness changes that occur during this time. Therefore, here we take an approach more similar to that presented in~\cite{kopp2009b}, taking into account the waiting time until the next mutation, during which the distribution of traits moves away from the optimum. We first seek the distribution of time steps before the next mutation establishes, $\tau(\Delta t)$. To do so, consider $F_z(\Delta t)$, which is the probability that for a fixed parental trait $z$, the time step before the next mutation is greater than $\Delta t$. This is given by an inhomogeneous Poisson process where the rate of new mutations that establish over the interval $[\Delta t, \Delta t + \epsilon]$ is:
\begin{equation*}
    Kf(z)\int_{z_K-z}^\infty \lambda(s)w(z-\sigma c \Delta t + s)ds,
\end{equation*}
where $\lambda(s)$ is again the distribution of mutational effect sizes, which we take to be Gaussian, and $w(z)$ is the probability of establishment. Note that we condition on mutations being large enough to establish as the new right-most trait by requiring $s>z_K-z$. Then $F_z(\Delta t)$ is the solution to the differential equation:
\begin{equation*}
    \frac{dF_z}{d\Delta t} = Kf(z)\int_{z_K-z}^\infty \lambda(s)w(z-\sigma c \Delta t + s)ds F_z(\Delta t).
\end{equation*}
Hence:
\begin{equation*}
    \ln F_z = Kf(z)\int_0^{\Delta t}\int_{z_K-z}^\infty \lambda(s)w(z-\sigma c t + s)ds dt.
\end{equation*}
To expand this to all parental traits $z$, we integrate over all traits:
\begin{equation*}
    \ln F = \int_{-\infty}^{z_K}\int_0^{\Delta t}\int_{z_K-z}^\infty Kf(z) \lambda(s)w(z-\sigma c t + s)ds dt dz.
\end{equation*}
In practice, we integrate only over the tip (from $z_{match}$ to $z_K$), as we expect the probability of large enough mutations arising from the bulk to be negligible. From this we can determine $\tau(\Delta t)$, which is given by:
\begin{equation*}
    \tau(\Delta t) = \frac{d(1-F)}{d\Delta t} = F(\Delta t)\int_{-\infty}^{z_K}\int_{z_K-z}^\infty Kf(z)\lambda(s)w(z-\sigma c \Delta t+s)dsdz. 
\end{equation*}

We define the distribution of mutation effect sizes that fix, conditioned on a waiting time of size $\Delta t$, to be:
\begin{equation*}
    \phi(s|\Delta t) = \frac{\int_{-\infty}^{z_K} Kf(z)\lambda(s)w(z-\sigma c \Delta t + s)dz}{\int_{-\infty}^{z_K}\int_{z_K-z}^\infty Kf(z) \lambda(s) w(z-\sigma c \Delta t+s)dsdz}.
\end{equation*}
Therefore the unconditional distribution of effect sizes given by:
\begin{align*}
    \phi(s) &= \int_0^\infty \frac{\int_{-\infty}^{z_K} Kf(z)\lambda(s)w(z-\sigma c \Delta t + s)dz}{\int_{-\infty}^{z_K}\int_{z_K-z}^\infty Kf(z) \lambda(s) w(z-\sigma c \Delta t+s)dsdz} \tau(\Delta t)d\Delta t \\
    &= \int_0^\infty \int_{-\infty}^{z_K} Kf(z)\lambda(s)w(z-\sigma c \Delta t + s) F(\Delta t)dzd\Delta t,
\end{align*}
and expected effect size is given by:
\begin{equation*}
    \bar{s} = \int_0^\infty \int_{-\infty}^{z_K} \int_{z_K-z}^\infty s K f(z)\lambda(s)w(z-\sigma c \Delta t + s) F(\Delta t)dsdzd\Delta t.
\end{equation*}
Due to the number of integrals, this calculation for $\bar{s}$ is computationally intensive. Computation for select parameter values show a value close to that found by the simplified calculation in Section~\ref{sec:first_sbar}, but further approximations of this equation would be necessary to compute $\bar{s}$ over the large range of parameters necessary to make predictions like those in Figures~\ref{fig:w1correction} and~\ref{fig:wcorrection}.

\section{Predicting typical phenotypic lineages} \label{sec:trajectories}

At the scale of the bulk, the transformation of Equation~\ref{eq:relative_pde} via $f(z)=K^{u(z)-1}$ yields a Hamilton-Jacobi equation (Eq.~\ref{eq:bulk_ode}). This class of equations can be equally represented as curve optimisation problems (see~\cite{calvez2020} and references therein). In the case of models of evolution such as ours, this dual representation provides further information on the ancestral lineages of the population~\citep{forien2022,champagnat2026}. The curves in question, which we will denote as $\gamma(t)$, are trajectories through phenotypic space, tracing possible ancestral lineages of an individual carrying trait $z$. The optimisation problem seeks to find the ancestral lineage which maximises fitness while minimising the mutational pathway necessary to follow the lineage. Analysis of the infinite population case has shown that the path of a typical ancestral lineage follows the predicted optimal phenotypic path~\citep{calvez2022,forien2022}. 

We start by writing the bulk equation (Eq.~\ref{eq:bulk_ode}) in the form:
\begin{equation}
    H(u') + cu' + R(z)=0, \label{eq:gen_HJ}
\end{equation}
where $R(z) = \frac{1}{\sigma_0}\left(1-\frac{z^2}{2}-\rho_K^0\right)$ is the scaled fitness of trait $z$, $\rho^0_K$ is the leading-order approximation of the total population density, and $H(p)=\frac{\sigma_0}{2}p^2$ is the Hamiltonian which describes mutations by a diffusion operator. Recall that $\sigma_0=\sigma\ln K$ (Sec.~\ref{sec:supp_bulk}). This has an associated Lagrangian, $L(v)=\frac{v^2}{2\sigma_0}$, which is related to the Hamiltonian by the Legendre transform:
\begin{equation*}
    H(p) = \sup_{v\in\bbR}(vp-L(v)).
\end{equation*}

The solution to Equation~\ref{eq:gen_HJ} (and equivalently, Equation~\ref{eq:bulk_ode}) in the infinite-population case can be written in the form of the curve optimisation problem:
\begin{equation*}
    u(z) = \sup_{\gamma(0)=z}\int^0_{-\infty}-L(\dot{\gamma}(t)+c)+R(\gamma(t))dt.
\end{equation*}
As with $c$, here $t$ is the rescaled unit of time. Unscaled, time is given by $\tilde{t}=\frac{t}{\sigma}$. We introduce a finite population constraint to this by requiring that for all time $t$, $u(\gamma(t))\geq0$~\citep{champagnat2026}. In this way, for any ancestral trait along the possible lineage $\gamma(t)$, the density of individuals carrying this trait is on the order of at least one individual. In other words, we require all ancestral traits to be expected in a finite population. Hence, the curve optimisation problem in the finite population case can be written formally as:
\begin{equation}
    u(z) = \sup_{\substack{\gamma(0)=z \\ \forall t,~u(\gamma(t))\geq 0}} \int_{-\infty}^0 -L(\dot{\gamma}(t)+c)+R(\gamma(t))dt. \label{eq:traj_int}
\end{equation}
The right-most trait $z_K$ can then be defined as the trait at which $u(z_K)=0$, which is the condition we apply to the original leading-order bulk approximation of the PDE problem (see Section~\ref{sec:supp_bulk}). Then a continuous curve that satisfies $u(\gamma(t))\geq0$ must satisfy $\gamma(t)\leq z_K$ for all time $t$.

We call the optimal trajectory $\Gamma(t)$, therefore:
\begin{equation*}
    u(z) = \int_{-\infty}^0-L(\dot{\Gamma}(t)+c)+R(\Gamma(t))dt.
\end{equation*}
Following the argument presented by~\cite{forien2022}, it is necessary that:
\begin{equation*}
    \lim_{t\to-\infty}\dot{\Gamma}(t)=0,
\end{equation*}
so that the fitness term $R(z)$ does not blow up. As $\Gamma(t)$ is the solution to the above optimisation problem, it is known that this curve will have special properties such that this limit implies it converges to a constant value. Maximising Equation~\ref{eq:traj_int} requires that $R(\Gamma(t))$ is maximised as $t\to-\infty$. Therefore:
\begin{equation*}
    \lim_{t\to-\infty}R(\Gamma(t))=\frac{1}{\sigma_0}\left(1-\min_{z\leq z_K}\frac{z^2}{2}-\rho_K^0\right),
\end{equation*}
so lineages converge to the fittest trait. In the infinite population case or when $z_K\geq0$ in a finite population, $\min_{z}\frac{z^2}{2}=0$, however in our finite population approximation, if the optimal trait has been lost ($z_K<0$), this corresponds to $\min_{z\leq z_K}\frac{z^2}{2}=\frac{z_K^2}{2}$. Finally, in order to ensure that $u(z)$ is finite, the integrand must converge to zero as $t\to-\infty$. Plugging in the above conditions on $\Gamma(t)$, this implies that:
\begin{equation*}
    -L(c)+\frac{1}{\sigma_0}\left(1 - \frac{z_K^2}{2}-\rho_K^0\right) = 0,
\end{equation*}
which only holds if:
\begin{equation*}
    \rho_K^0 = 1 - \frac{c^2}{2}-\min_{z\leq z_K}\frac{z^2}{2}.
\end{equation*}
Hence, we find a transition. When $z_K\geq0$, the ancestral lineage $\Gamma(t)$ converges to 0 as $t\to-\infty$, behaving as in the infinite population case. Since this is true regardless of the trait $z$ of the current-day individual, this implies that the ancestral lineages of individuals in the population should converge to a common ancestor carrying the optimal trait. This prediction has been verified by analysis of the stochastic processes under the infinite-population assumption~\citep{calvez2022,forien2022}. However, when $z_K<0$, the deterministic prediction is that the ancestral lineages converge to a common ancestor carrying trait $z_K$ instead. In both cases, the lineages converge to a common ancestor with the highest fitness. The fitness of the population depends always on the fitness of this fittest individual, $\min_{z\leq z_K}\frac{z^2}{2}$. When that individual is optimal, there is no additional fitness cost, otherwise an extra load is incurred: the loss load. 

We can further derive an equation for the trajectory of the ancestral lineage. This has already been done in the infinite population case~\citep{forien2022}, and this solution holds for the finite population case when $z_K\geq0$. Here, we focus on $z_K<0$. By the Euler-Lagrange optimality condition, it can be shown that the trajectory optimising Equation~\ref{eq:traj_int} solves the ODE:
\begin{equation*}
    \ddot{\Gamma}=\Gamma,
\end{equation*}
the solution to which is given by:
\begin{equation*}
    \Gamma(t) = Ae^t + Be^{-t},
\end{equation*}
for which $A$ and $B$ are unknown constants. To satisfy the condition that $\Gamma(t)$ converges to $z_K$ (assuming $z_K<0$), it must be true that $\Gamma(t)$ is a trajectory that hits $z_K$ at some finite time $T$, then remains constant past this time. On the interval $t\in[T,0]$, then, $\Gamma(t)$ is the solution to the above ODE, subject to three conditions. First, that $\Gamma(0)=z$ and $\Gamma(T)=z_K$. Additionally, by standard elliptic regularity theory $\ddot{\Gamma}(t)$ is square-integrable, and so $\dot{\Gamma}(t)$ is continuous~\citep[Lemma 7.2]{calvez2022b}. Therefore, $\Gamma(t)$ must satisfy the condition that $\dot{\Gamma}(T)=0$. Altogether, this leaves us with three unknowns, $A$, $B$, and $T$, and three constraints: $\Gamma(0)=z$, $\Gamma(T)=z_K$, and $\dot{\Gamma}(T)=0$. From this we find that the lineage reaches $z_K$ at time:
\begin{equation*}
    T = \textrm{arccosh}\left(\frac{z}{z_K}\right).
\end{equation*}
Recall the scaling of time. Unscaled, this becomes:
\begin{equation*}
    \tilde{T} = \frac{1}{\sigma}\textrm{arccosh}\left(\frac{z}{z_K}\right).
\end{equation*}
So as $\sigma$ increases, the time to reach $z_K$ decreases. This time also increases as $z_K$ approaches 0, which is true as $c$ decreases or $K$ increases. 

Before reaching $z_K$, or on $t\in[T,0]$, the trajectory of this lineage is given by:
\begin{equation*}
    \Gamma(t) = -\frac{1}{2}\left(z+\sqrt{z^2-z_K^2}\right)e^t+\frac{1}{2}\left(3z+\sqrt{z^2-z_K^2}\right)e^{-t}.
\end{equation*}
Allowing for the case when $z_K\geq0$ and the trajectory follows that predicted by the infinite population case, the typical trajectory for a lineage of an individual carrying trait $z$, given the right-most trait of the population is $z_K$, is:
\begin{equation}
    \Gamma(t) = \begin{cases}
        ze^{-t} & z_K \geq 0 \\
        \begin{cases}
        -\frac{1}{2}\left(z+\sqrt{z^2-z_K^2}\right)e^t+\frac{1}{2}\left(3z+\sqrt{z^2-z_K^2}\right)e^{-t} & t\in [T,0] \\
        z_K &  t \in (-\infty,T)
        \end{cases} & z_K<0
    \end{cases}. \label{eq:typ_lineage}
\end{equation}
This result holds for small, but frequent mutations ($\eta\simeq1$ and $\sigma\ll1$), in the regime of pervasive clonal interference~\citep{neher2013,roques2017}. Although we do not have an analytical prediction of $z_K$, we can compare our prediction to simulations by plugging in the mean value of observed $Z_K$ over time. Figure~\ref{fig:lineages} shows that this trajectory matches well with the stochastic lineages. 

Leaving the regime of pervasive clonal interference, dynamics change (Fig.~\ref{fig:mu_lineages}) and the lineage cannot be predicted deterministically. When $\sigma_{mut}^2\ll9\eta$ (Fig.~\ref{fig:mu_lineages}, left two panels), we find ourselves in the Gaussian regime~\citep{burger2000:ch4}. Here, the mean population fitness can be well-predicted by the Gaussian allelic approximation of the genetic variance and a Gaussian fitness function (\cite{burger1999}; see also Fig.~\ref{fig:burger_sim}, left panel). As the mutational variance is larger than in the regime of pervasive clonal interference, adaptation is no longer limited by the tip and larger mutations can retrieve the optimal trait more easily. As mutations grow rarer and larger ($\sigma_{mut}^2\gg20\eta$), the population will enter into the house-of-cards regime, where the mutational variance is larger than the genetic variance~\citep{burger2000:ch4}. Again, individuals from the bulk can produce offspring past the tip, but additionally the probability of a mutant arising past the tip becomes essentially independent of its origins and the assumption of a normal phenotypic distribution breaks down. Simulations show that in this case, the population derives itself from a lucky lineage that undergoes regular large mutations that overshoot the optimal trait before drifting back behind the optimal trait (Fig.~\ref{fig:mu_lineages}, right panel). Finally, on the opposite end of the spectrum from our pervasive clonal interference regime is the adaptive walk, where mutations are so rare that adaptation occurs as a series of selective sweeps~\citep{kopp2007,kopp2009a,kopp2009b}.

\section{Connections to stochastic branching models} \label{sec:stoch_models}

Our work has sought to approximate the dynamics of a stochastic process in a deterministic framework. The cut-off method we use has been more formally linked to rigorous stochastic analyses in other models. In particular, the predictions of this approach applied to the F-KPP equation~\citep{brunet1997,brunet2001} were validated with formal analysis of the corresponding stochastic differential equation by \cite{mueller2011}. More recently, \cite{roberts2021} analysed a branching Brownian motion process similar to models of rapid asexual evolution~\citep{neher2013} and rigorously retrieved results matching those predicted by a cut-off analysis. While the stochastic model we approximate here has yet to be analysed, previous work on related models support our methods and results. Here, we will outline the connections between our work and existing work in the probability theory literature.

\subsection{Connections to constrained Hamilton-Jacobi equations} \label{sec:champ}

A similar model to ours, which does not incorporate density-dependent effects or environmental change, has been recently studied by \cite{champagnat2026}. Here, individuals are modelled as particles undergoing a branching process with births, deaths, and mutations. They scaled the density of particles by the value $K$, and found that in the limit as $K\to \infty$, they could derive a constrained Hamilton-Jacobi equation which described the system. This serves to motivate our use of the change of variables $K^{u(t)-1}$, as well as the the use of a cut-off, as we explain here.

To make the connection with our work more concrete, here we will loosely describe how these results apply to the model most similar to our model. Consider a population of individuals subject to a birth rate of 1 and a death rate of $\frac{x^2}{2}$, with no density dependence. Mutations occur with every birth according to a Gaussian distribution with mean 0 and variance $\left (\frac{1}{\ln K}\right )^2$. The integro-differential equation describing this system in a deterministic setting is given by:
\begin{equation*}
    \partial_tf(t,x)=\frac{1}{\sigma}\int_\bbR G\left(\frac{x-x'}{\sigma}\right)f(t,x')dx' - \frac{x^2}{2}f(t,x),
\end{equation*}
where $G(y)$ is the Gaussian distribution with mean 0 and variance 1, and $f(t,x)$ is the density of the population at time $t\geq 0$ carrying trait $x\in\bbR$. Applying the change of variables $f(t,x)=K^{u(t,x)}$ and taking $\mathcal{O}(1)$ terms yields the Hamilton-Jacobi equation in $u(t)$:
\begin{equation}
    \partial_tu(t,x) = H(\partial_xu)+R(x), \label{eq:u0_PDE}
\end{equation}
where:
\begin{equation*}
    H(p)=\int_\bbR (e^{p y}-1)G(y)dy,
\end{equation*}
is the Hamiltonian and:
\begin{equation*}
    R(x) = 1 - \frac{x^2}{2},
\end{equation*}
is the net growth rate of an individual with trait $x$.

In a stochastic setting, we imagine a finite number of particles evolving according to the same birth rate, death rate, and mutational distribution. This system is initialised with an inhomogeneous Poisson distribution with rate $K^{\beta_K(0,x)}$, where in the limit $K\to\infty$, $\beta_K(0,x)$ will converge to some function $\beta(0,x)$. In this way, $K^{\beta_K(0,x)}$ describes the initial trait distribution of the population. 

The aim is to understand the number of particles in a neighbourhood of size $\delta$ around some trait $x$, denoted $N_t^{A^{x,\delta}}$. This is scaled by $K$, such that:
\begin{equation*}
    N_t^{A^{x,\delta}} = K^{\beta^\delta_K(t,x)}.
\end{equation*}
Taking the limit of $K\to\infty$ and shrinking the neighbourhood around $x$ with $\delta\to0$, then under the appropriate conditions $\beta^\delta_K(t,x)$ will converge in probability to: 
\begin{equation*}
    u(t,x)=\sup_{\substack{\gamma(t)=x \\ \forall s\in [0,t], \ u(s,\gamma(s))\geq 0}} \beta(0,\gamma(0))+\int_0^t -L(\dot{\gamma}(s))+R(\gamma(s)),
\end{equation*}
where $L(v)$ is the Lagrangian associated with $H(p)$ and $\gamma(s)$ must be absolutely continuous over the interval $[0,t]$ \cite[Theorem 1.3 and Remark 1.6]{champagnat2026}. Further, up to perturbing slightly the initial condition, $u(t,x)$ is the solution to the Hamilton-Jacobi equation \cite[Theorem 1.4]{champagnat2026}:
\begin{equation}
    \begin{cases}
        \partial_tu(t,x) = H(\partial_xu)+R(x), & (t,x)\in\Omega \\
        u(t,x)=0 & (t,x)\in\partial\Omega, \ t>0 \\
        u(0,x)=\beta(0,x) & \forall x, \ \beta(0,x)>0
    \end{cases}, \label{eq:state_constrained_u0}
\end{equation}
where:
\begin{equation*}
    \Omega=\{(t,x)\in\tilde{\Omega} | u(t,x)>0\},
\end{equation*}
requires that the density of individuals carrying trait $x$ at time $t$ is of at least order 1, and
\begin{equation*}
    \tilde{\Omega}=\{(t,x)|\ \exists \gamma(t)=x, \ \forall s\in[0,t], \ \beta(0,\gamma(0))+\int_0^t -L(\dot{\gamma}(s))+R(\gamma(s)) \geq 0\},
\end{equation*}
requires that there exists some lineage $\gamma(s)$ to $x$ in which all ancestral traits of the trait $x$ existed (that is, are of a density order 1). 

Equation~\ref{eq:state_constrained_u0} is exactly Equation~\ref{eq:u0_PDE} constrained to values of order 1, thus deriving a rigorous connection between the stochastic system and the deterministic equation. This is the same scaling and cut-off approach that we use in Section~\ref{sec:supp_bulk}. Further, the restriction of $u(s,\gamma(s))\geq 0$ used to define $u(t,x)$ is the same restriction applied in Equation~\ref{eq:traj_int} in order to define the typical ancestral lineage. It's important to note, however, that the above describes model behaviour over short timescales. Our work has focused on behaviour on long timescales, when a stationary distribution has been reached. Note also that the above considers only the scale of the bulk, likely a consequence of taking the limit $K\to\infty$. We see in Equation~\ref{eq:match_point} that in this limit, the interval of traits over which the linearised tip must be considered disappears. Understanding the relationship between short timescales and long timescales, as well as the role of the behaviour at the tip will require more investigation into the stochastic models. 

\subsection{Appearance of the Airy solution in a branching Brownian motion model}

At the tip, the linearisation of the mortality term and the assumption that the population is behind the optimal trait yields a model that follows similar rules to that of a mutation accumulation model at the tip. In fact, through a change of variables, the results of~\cite{roberts2021} can be applied to our model on the scale of the tip. Here, we will show how our solution for the density of individuals at the tip (Eq.~\ref{eq:tip_solution}) can be retrieved in their results.

We consider the tip as a collection of particles $X_{i,n}(t)$ which mutate by a Brownian motion with mutational variance $\sigma_n^2$. The index $i$ represents the particles in the system, while the index $n$ separates different particle systems with different variances. Each particle is described by its distance from $z_K$, that is $x=z-z_K$. These particles branch with births at rate 1 and die at rate $\rho+\frac{z_K^2}{2}+z_Kx$. Additionally, environmental change is modelled by a drift term which pushes the particles back at rate $\sigma_nc$. Here we assume that $\lim_{n\to\infty}\sigma_n=0$. Our goal is to understand the limiting distribution describing the system of particles as $n\to\infty$ and the mutational variance tends to zero. Note that as this is a model of branching Brownian motion, it is not the same as what we simulate in our stochastic simulations (Sec.~\ref{sec:simulations}), but the stochastic analogue of the PDE with diffusion. As we have previously noted, however, for small mutational variance and high mutational rate, simulations with diffusion and simulations with mutations drawn at birth from a Gaussian distribution show the same behaviour (Fig.~\ref{fig:mut_rates}).

To apply the results of \cite{roberts2021}, we require a change of variables:
\begin{equation*}
    Y_{i,n}(t)=a_nX_{i,n}(b_nt)-d_n,
\end{equation*}
where:
\begin{equation*}
    a_n = \frac{c}{\sigma_n\lambda_n}, \ \ \ b_n = \frac{\lambda_n^2}{c^2}, \ \ \ d_n = -\frac{c}{|z_K|\sigma_n\lambda_n}\left(1-\rho-\frac{z_K^2}{2}\right),
\end{equation*}
where $\lambda_n$ is a arbitrary sequence converging to zero. Under this change of variables, the system of particles $Y_{i,n}(t)$ undergo a Brownian motion with mutational variance 1. They branch with a birth rate $b_n = \frac{\lambda_n^2}{c^2}$ and die at rate $d_n(y)=\frac{\lambda_n^2}{c^2}\left(1+|z_K|\frac{\sigma_n\lambda_n}{c}y\right)$. Therefore the net rate is $b_n-d_n(y)=\beta_ny$, where $\beta_n=|z_K|\sigma_n\frac{\lambda_n^3}{c^3}$. Note that:
\begin{equation*}
    \lim_{n\to\infty}\frac{\lambda_n^3}{\beta_n}=\lim_{n\to\infty} \frac{c^3}{|z_K|\sigma_n}=\infty,
\end{equation*}
satisfying Assumption 1.2 in \cite{roberts2021}. 

Equation~1.6 predicts the particles $X_{i,n}(t)$ will generally be to the left of 0, that is, that individuals will generally carry traits left of $z_K$, consistent with our assumption. Next, we will assume good initial conditions \cite[Equations 1.9, 1.10]{roberts2021}, which require enough particles near the right edge (that is, enough individuals are near the right-most trait). Additionally, we let $\rho=1-\frac{c^2}{2}-\frac{z_K^2}{2}+\frac{\xi_0}{2}|2z_K\sigma_n|^{2/3}$, the predicted equilibrium density. Then we consider the following random probability measure:
\begin{equation*}
    \frac{\sum_{i=1}^{N_n(t)}\exp\left(\frac{c}{\sigma_n}X_{i,n}(t_n)\right)\delta_{-\left(\frac{2|z_K|}{\sigma_n^2}\right)^{1/3}X_{i,n}(t_n)}}{\sum_{i=1}^{N_n(t)}\exp\left(\frac{c}{\sigma_n}X_{i,n}(t_n)\right)},
\end{equation*}
where $\delta_x$ is a Dirac measure.
This amounts to weighting each particle by $\exp\left(\frac{c}{\sigma_n}x\right)$ when it is counted.
Subject to the correct bounding on time \cite[Equation 1.13]{roberts2021}, this will converge in distribution as $n\to\infty$ to the probability measure associated the following probability density function~\cite[Theorem 1.2]{roberts2021}:
\begin{equation*}
    h(y)=\frac{Ai(\xi_0+y)}{\int_0^\infty Ai(\xi_0+z)dz}.
\end{equation*}
That is, for any test function $g(y):\bbR\to\bbR$:
\begin{equation*}
    \lim_{n\to\infty} \frac{\sum_{i=1}^{N_n(t)}\exp\left(\frac{c}{\sigma_n}X_{i,n}(t_n)\right)g\left({-\left(\frac{2|z_K|}{\sigma_n^2}\right)^{1/3}X_{i,n}(t_n)}\right)}{\sum_{i=1}^{N_n(t)}\exp\left(\frac{c}{\sigma_n}X_{i,n}(t_n)\right)} =\frac{\int_{-\infty}^0g\left(y\right)Ai\left(\xi_0+y\right)dx}{\int_{-\infty}^0Ai\left(\xi_0+y\right)dx}.
\end{equation*}

Loosely speaking, this means that for some collection of particles $X_{i,n}$ moving with small variance $\sigma_n$, the weighted mass of a particle at distance $x$ from $z_K$ may be measured approximately by $Ai\left(\xi_0-\left(\frac{2|z_K|}{\sigma^2}\right)^{1/3}x\right)$. Counteracting this weight, then, requires multiplying by $\exp\left(-\frac{c}{\sigma_n}x\right)$. Thus, we see the tip solution (Eq.~\ref{eq:tip_solution}) emerge as an approximative description of the density of particles $X_{i,n}$ for small $\sigma_n$. We also see the scaling of $\sigma^{2/3}$ naturally appear in the division of $x$ by this value.  Unfortunately, this connection can give us no information about the value of the unknown constant $A$ in our tip solution.



\end{document}